\documentclass[a4paper,11pt]{article}

\usepackage{jheppub} 
\usepackage[T1]{fontenc} 

\usepackage{graphicx,multirow,subfigure}
\usepackage{float}
\usepackage{bm}
\usepackage{amsmath}
\usepackage{amssymb}
\usepackage{amscd}
\usepackage{latexsym}
\usepackage{slashed}
\usepackage{color}
\usepackage{graphicx}
\usepackage{color}
\usepackage{verbatim}
\usepackage{rotating}
\usepackage{diagbox}

\def\bea{\begin{eqnarray}}
\def\eea{\end{eqnarray}}

\title{ \centering 
Extra Higgs Boson and $Z'$ as  Portals\\
to Signatures of  Heavy Neutrinos at the LHC
}

\author[a]{Elena Accomando,}
\author[a,b]{Luigi Delle Rose,}
\author[a,b]{Stefano Moretti,} 
\author[b]{Emmanuel Olaiya,}
\author[b]{Claire H. Shepherd-Themistocleous}

\emailAdd{E.Accomando@soton.ac.uk}
\emailAdd{L.Delle-Rose@soton.ac.uk}
\emailAdd{S.Moretti@soton.ac.uk}
\emailAdd{Emmanuel.Olaiya@stfc.ac.uk}
\emailAdd{Claire.Shepherd@stfc.ac.uk}

\affiliation[a]{School of Physics and Astronomy, University of Southampton, Highfield, Southampton SO17 1BJ, United Kingdom}
\affiliation[b]{Particle Physics Department, Rutherford Appleton Laboratory, Chilton, Didcot, Oxon OX11 0QX, United Kingdom}

\abstract{\noindent
In this paper, we discuss the potential of observing heavy neutrino ($\nu_h$) signatures of a $U(1)_{B-L}$ 
enlarged Standard Model (SM) encompassing three heavy Majorana
neutrinos alongside the known light neutrino states at the Large Hadron Collider (LHC). We exploit the theoretical decay via a pair of heavy (non-SM-like)
Higgs boson and $Z'$ production followed by  $\nu_h \rightarrow l^\pm W^{\mp (*)}$ and $\nu_h \rightarrow \nu_l Z^{(*)}$ decays, ultimately yielding 
a $3l+2j+E_{T}^{\rm{miss}}$ signature and,  depending upon how boosted the final state objects are, we define different possible selections aimed at improving the signal to background ratio in LHC Run 2 data for a wide range of heavy neutrino masses. 
}

\begin{document}
\maketitle
\flushbottom

\section{Introduction}
\label{sec:introduction}
Models with an additional Abelian group, $U(1)'$ , with respect to the Standard Model (SM) symmetry can have their origin in a Grand Unification Theory (GUT) and it is possible for this extra symmetry  to be  broken at energies accessible at the CERN Large Hadron Collider (LHC). 
There are several realisations of GUTs that embed this dynamics, such as, e.g., $E_6$, String Theory motivated, $SO(10)$ and Left-Right (LR) symmetric models \cite{Langacker:1980js,Hewett:1988xc,Faraggi:1990ita,Faraggi:2015iaa,Faraggi:2016xnm,Randall:1999ee,Accomando:2010fz}. 

An interesting phenomelogical case is the one where the conserved charge of the extra Abelian symmetry is  $B-L$, with $B$ and $L$ the baryon and lepton numbers, respectively, containing three heavy Right-Handed (RH) neutrinos, one extra heavy neutral gauge boson, $Z'$, and an additional Higgs boson responsible for the $U(1)_{B-L}$ symmetry breaking. Herein, the cancellation of the $U(1)_{B-L}$ gauge and gravitational anomalies naturally predicts the RH neutrinos that can be at the TeV scale, thus realising a low-scale seesaw mechanism for neutrino mass generation. Collider signatures emerging from such states are
the concern of this paper. 

While the signatures connected to the $Z'$ and an additional Higgs boson have repeatedly been
studied in  specialised literature \cite{Khalil:2007dr,Basso:2008iv,Basso:2010hk,Basso:2010pe,Basso:2010yz,Basso:2010jm,Accomando:2010fz,Basso:2011na,Basso:2012sz,Basso:2012ux,Accomando:2013sfa,Accomando:2015cfa,Accomando:2015ava, Okada:2016gsh}, the signatures emerging from the heavy neutrino sector have seen less investigation. From the diagonalisation of the neutrino mass matrix one obtains three very light, mostly Left-Handed (LH), neutrinos ($\nu_l$), which are identified as the SM ones, and three heavy, mostly Right-Handed (RH), neutrinos ($\nu_h$), with  very small mixing with the light $\nu_l$'s, thereby obtaining very small yet non-vanishing couplings to the gauge bosons. Moreover, owing to the mixing in the scalar sector, the Yukawa interaction of the heavy neutrinos with the heavier Higgs, $H_2$, also provides the coupling of the $\nu_h$'s to the SM-like Higgs boson, $H_1$. These non-zero couplings enable, in particular, the $gg\to H_1\to \nu_h\nu_h$ production mode if $m_{\nu_h} < m_{H_1}/2$ and the subsequent $\nu_h \rightarrow l^\pm W^{\mp *}$ and  $\nu_h \rightarrow \nu_l Z^{*}$ (off-shell) decays \cite{Brooijmans:2012yi,Caputo:2017pit}, an alternative to the case of SM-like Higgs decays instead into one light and one heavy neutrino    
 \cite{Gago:2015vma,BhupalDev:2012zg,Cely:2012bz,Shoemaker:2010fg,Antusch:2016vyf}.

A particular feature of such a heavy neutrino pair signature is that  decay width of the heavy neutrino is small and its lifetime large, so that it turns out to be  a long-lived particle and, over a large portion of the $U(1)_{B-L}$ parameter space, its lifetime can be such that it can decay inside the LHC detectors, thereby producing a very distinctive signature with Displaced Vertices (DVs). This is due to the fact that the 
 heavy neutrino couplings to the weak gauge bosons ($W^\pm$ and $Z$) are
proportional to the ratio of light over heavy neutrino masses, which is extremely small in a type-I seesaw scenario.  Ref. \cite{Accomando:2016rpc} proved the feasibility of establishing
this signal at the LHC during Run 2, primarily because of a negligibly small background contribution\footnote{An experimentally resolvable non-zero lifetime along with a mass determination for the heavy neutrino would potentially also enable an indirect measurement of the light neutrino mass, as remarked in Ref.~\cite{Basso:2008iv}.}. 

When the $\nu_h$ mass grows larger than $m_{H_1}/2$, the aforementioned production mode is no longer available, as the SM-like Higgs become immediately off-shell, owing to its small width, of 5 MeV or so. At the same time, simply because of the phase space enhancement entering the total width of the $\nu_h$ state, the latter stops producing DVs. Hence, one is forced to search for decay products which stem directly  from the interaction point. In these condictions, the presence of the background becomes sizable. Furthermore, to stay with the aforementioned decay patterns, i.e.,   $\nu_h \rightarrow l^\pm W^{\mp}$ and $\nu_h \rightarrow \nu_l Z$, wherein the weak gauge bosons can be on-shell, one also realises that a significant boost can be given to the $W^\pm$ and $Z$ decay products, so that they become more collimated in phase space as $m_{\nu_h}$ grows larger.   

It is the purpose of this paper to examine heavy neutrino pair production and decay  at the LHC, specifically for the
signature  $3l+2j+E_{T}^{\rm{miss}}$ (where $l$ represents an electron or muon, $j$ a jet and $E_{T}^{\rm{miss}}$ the missing transverse energy), when their mass
is the range $m_{\nu_h}>m_{H_1}/2$ up to the TeV scale. We shall see that two production mechanism will turn out to be useful, depending on the actual
value of $m_{\nu_h}$, namely, $gg\to H_2\to \nu_h\nu_h$ (for intermediate mass values in the range $m_{H_1}/2<m_{\nu_h}<m_{H_2}/2$)
and $q\bar q\to Z'\to \nu_h\nu_h$ (for heavy mass values in the range $m_{H_2}/2<m_{\nu_h}<M_{Z'}/2$). Notice in fact that the $H_2$ mass can be accessible at the LHC up to
several hundreds of GeV while current limits on the $Z'$ one are a few TeV \cite{Accomando:2016upc,Accomando:2016eom,Accomando:2016sge}.

This paper is organised as follows. Sect.~\ref{sec:model} reviews
the model under study together with an overview of its allowed parameter
space. Then we describe the relevant production and decay processes of the heavy neutrino states in different sections depending on their mass, including
presenting our numerical results. We then conclude in 
Sect.~\ref{sec:summa}.

\section{A Minimal Abelian Extension of the SM}
\label{sec:model}
In this paper, we deal with a minimal renormalisable Abelian extension of the SM with  only the  matter content necessary to satisfy the cancellation of all gauge and  gravitational anomalies. Hence, we augment each of the three lepton families by a RH neutrino which is a singlet under the SM gauge group with $B-L$ = $-1$ charge. Furthermore, 
in the scalar sector, we introduce a complex scalar field $\chi$, besides the SM-like Higgs doublet $H$, to trigger spontaneous symmetry breaking of the extra Abelian gauge group. The new scalar $\chi$ has $B-L$ = 2 charge and is a SM singlet. Its Vacuum Expectation Value (VEV), $x$, gives mass to the new heavy neutral gauge boson $Z'$ and provides the Majorana mass to the RH neutrinos through a Yukawa coupling. (It is the latter that dynamically implements the type-I seesaw mechanism.) 

The presence of two Abelian gauge groups allows for a gauge invariant kinetic mixing operator of the corresponding Abelian field strengths. This mixing is normally removed from the kinetic Lagrangian through a suitable transformation (rotation and rescaling), thus restoring its canonical form. However, it is reintroduced, through a coupling, $\tilde g$, in the gauge covariant derivative which thus acquires a non-diagonal structure
\bea
\label{eq:gaugecovder}
\mathcal D_\mu = \partial_\mu + \ldots + i g_1 Y B_\mu + i \left( \tilde g Y + g'_1 Y_{B-L} \right) B_\mu',
\eea
where $Y$ and $Y_{B-L}$ are  the hypercharge and the $B-L$ quantum numbers, respectively, while $B_\mu$ and $B'_\mu$ are the corresponding Abelian fields. Other parameterisations, with a non-canonical diagonalised kinetic Lagrangian and a diagonal covariant derivative, are, however, completely equivalent.
The details of the kinetic mixing and its relation to the $Z-Z'$ mixing, which we omit in this work, can be found in \cite{Basso:2008iv,Coriano:2015sea,Accomando:2016sge}. 

The $Z-Z'$ mixing angle in the neutral gauge sector is strongly bounded indirectly by the EW Precision Tests (EWPTs) and directly by the LHC data \cite{Langacker:2008yv,Erler:2009jh,Cacciapaglia:2006pk,Salvioni:2009mt,Accomando:2016sge} to small values, i.e, $|\theta'| \lesssim 10^{-3}$. In the $B-L$ model under study, we find
\bea
\label{eq:thetapexpandend}
\theta ' \simeq \tilde g \frac{M_Z \, v/2}{M_{Z'}^2 - M_Z^2} \,,
\eea
where $v$ is the VEV of the SM-like Higgs doublet, $H$. In this case, the bound on the $Z-Z'$ mixing angle can be satisfied by either $\tilde g \ll 1$ or $M_Z / M_{Z'} \ll 1$, the latter allowing for a generous interval of allowed values for $\tilde g$.

After spontaneous symmetry breaking, two mass eigenstates, $H_{1,2}$, with masses $m_{H_{1,2}}$, are obtained from the orthogonal transformation of the neutral components of $H$ and $\chi$. The mixing angle of the two scalars is denoted by $\alpha$. Moreover, we choose $m_{H_1} \le m_{H_2}$ and we identify $H_1$ with the $125$ GeV SM-like Higgs discovered at the LHC.
The couplings between the light (heavy) scalar and the SM particles are equal to the SM rescaled by $\cos \alpha$ ($\sin \alpha$). The interaction of the light (heavy) scalar with the extra states introduced by the Abelian extension, namely the $Z'$ and the heavy neutrinos, is instead controlled by the complementary angle $\sin \alpha$ ($\cos \alpha$).

Finally, the Yukawa Lagrangian is
\bea
\mathcal L_Y =  \mathcal L_Y^{SM}  - Y_\nu^{ij} \, \overline{L^i} \, \tilde H \, {\nu_R^j}   - Y_N^{ij} \, \overline{(\nu_R^i)^c} \, {\nu_R^j} \, \chi + \, {\rm h.c.}, 
\eea
where $\mathcal L_Y^{SM}$ is the SM contribution. The Dirac mass, $m_D = 1/\sqrt{2}\, v Y_\nu$, and the Majorana mass for the RH neutrinos, $M = \sqrt{2}\, x Y_N$, are dynamically generated through the spontaneous symmetry breaking and, therefore, the type-I seesaw mechanism is automatically realised. Notice that $M$ can always be taken real and diagonal without loss of generality. For $M \gg m_D$, the masses of the physical eigenstates, the light and the heavy neutrinos, are, respectively, given by $m_{\nu_l} \simeq - m_D^T  M^{-1} m_D$ and $m_{\nu_h} \simeq M$. The light neutrinos are dominated by the LH SM components with a very small contamination of the RH neutrinos, while the heavier ones are mostly RH. The contribution of the RH components to the light states is proportional to the ratio of the Dirac and Majorana masses. After rotation into the mass eigenstates the charged and neutral currents interactions involving one heavy neutrino are given by
\bea
\mathcal L = \frac{g_2}{\sqrt{2}} \, V_{\alpha i} \, \bar l_\alpha \gamma^\mu P_L \nu_{h_i} \, W^-_\mu + \frac{g_Z}{2 \cos \theta_W} V_{\alpha \beta} V_{\alpha i}^* \, \bar \nu_{h_i} \gamma^\mu P_L \nu_{l_\beta} \, Z_\mu
\eea
where $\alpha, \beta = 1,2,3$ for the light neutrino components and $i = 1,2,3$ for the heavy ones. The sum over repeated indices is understood. In particular, $V_{\alpha \beta}$ corresponds to the Pontecorvo-Maki-Nakagawa-Sakata (PMNS) matrix while $V_{\alpha i}$ describes the suppressed mixing between light and heavy states. Notice also that the $Z \nu_h \nu_h$ vertex is $\sim V_{\alpha i}^2$ and, therefore, highly dumped. These interactions are typical of a type-I seesaw extension of the SM. The existence of a scalar field generating the Majorana mass for RH neutrinos through a Yukawa coupling, which is a characteristic feature of the Abelian extensions of the SM, allows for a new and interesting possibility of producing a heavy neutrino pair from the SM-like Higgs (besides the obvious heavy Higgs mode). The corresponding interaction Lagrangian is given by 
\bea
\mathcal L = - \frac{1}{\sqrt{2}}Y_N^{k} \sin \alpha \, H_1 \, \bar \nu_{h_k} \nu_{h_k} = - g'_1 \frac{m_{\nu_{h,k}}}{M_{Z'}}  \sin \alpha \, H_1 \, \bar \nu_{h_k} \nu_{h_k},
\eea
where, in the last equation, we have used $x \simeq M_{Z'}/(2 g'_1)$. This expression for the VEV of $H_2$, $x$, neglects the sub-leading part that is proportional to $\tilde g$. For our purposes, this approximation can be safely adopted \cite{Basso:2010jm,Accomando:2016rpc}. The interaction between the light SM-like Higgs and the heavy neutrinos is not suppressed by the mixing angle $V_{\alpha i}$ but is controlled by the Yukawa coupling $Y_N$ and  scalar mixing angle $\alpha$. 

For illustrative purposes we assume that the PMNS matrix is equal to the identity matrix and that both neutrino masses, light and heavy, are degenerate in  flavour. In this case the elements of the neutrino mixing matrix $V_{\alpha i}$ are simply given by $m_D/M \simeq \sqrt{m_{\nu_l}/m_{\nu_h}}$.

\section{Heavy neutrino properties}

\begin{figure}[t]
\centering
\subfigure[]{\includegraphics[scale=0.7]{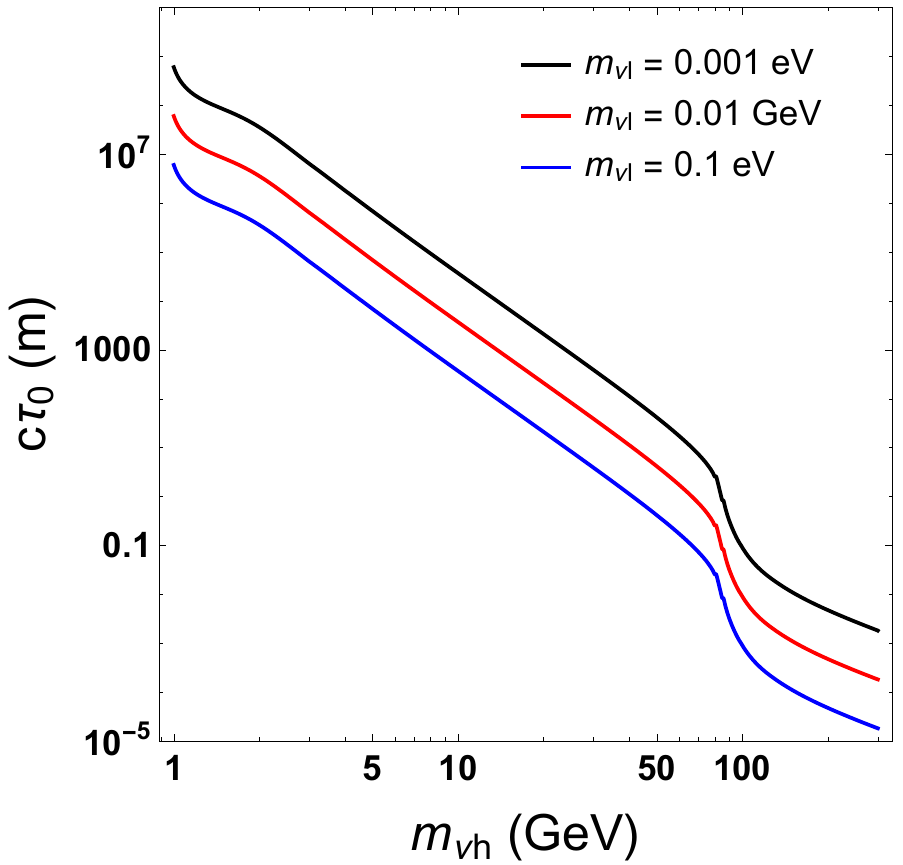}}
\subfigure[]{\includegraphics[scale=0.68]{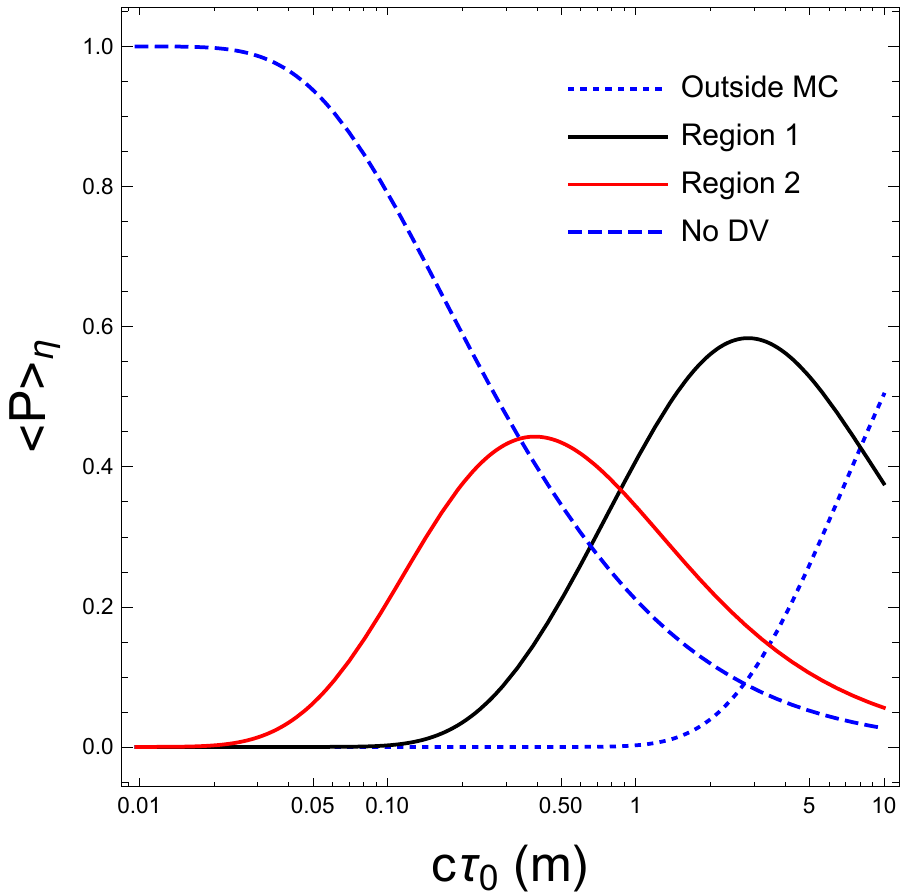}}
\caption{(a) Proper decay length of the heavy neutrino as a function of the heavy neutrino mass 
$m_{\nu_h}$. (b) Averaged decay probability of the heavy neutrino for $\beta \gamma = 1$ in 
different regions of the detector (see text).}
\label{fig:decaylength}
\end{figure}

Heavy neutrinos are characterised by a proper decay length that depends on the mass of the heavy neutrino itself and on the mass of the light neutrinos. This property has been discussed in details in Ref.~\cite{Basso:2008iv,Accomando:2016rpc}. Here, we briefly summarise the result. In Fig.~\ref{fig:decaylength} (a), we plot the heavy neutrino's proper decay length as a function of the neutrino's mass for three different values of the light neutrinos mass. The blue curve corresponds to the upper limit on the light neutrino mass. As one can see, the proper decay length decreases with increasing the neutrinos mass, ranging from 10$^9$ meters and more to the sub-millimeter. One can roughly define $m_{\nu h}\simeq M_W$ as a mass threshold for the heavy neutrino, dividing the short-lived from the long-lived regime. The difference between the two regimes is reflected in the detection strategy of such a particle. In Fig.~\ref{fig:decaylength} (b), we show the decay probability of the heavy neutrino in the various regions of the detector, averaged on the pseudo-rapidity $\eta$ at which the heavy neutrino is emitted off the parent particle, as a function of the proper decay length. The blue dashed line corresponds to the case when the heavy neutrino is detected like all other short-lived particles, i.e. no displaced vertices are supposed to be produced. The red and black solid lines refer to the two cases when the heavy neutrino decays and generates a displaced vertex in the inner tracker and in the muon chamber, respectively. The dotted blue curve is when the neutrino decays outside the muon chamber, that is the heavy neutrino is undetected and gives rise to missing transverse energy. We refer to Ref.~\cite{Accomando:2016rpc} for the long-lived heavy neutrino search via displaced vertices techniques while in Section 4, a study for detecting and reconstructing short-lived heavy neutrinos will be presented. 

\begin{figure}[t]
\centering
\includegraphics[scale=0.5]{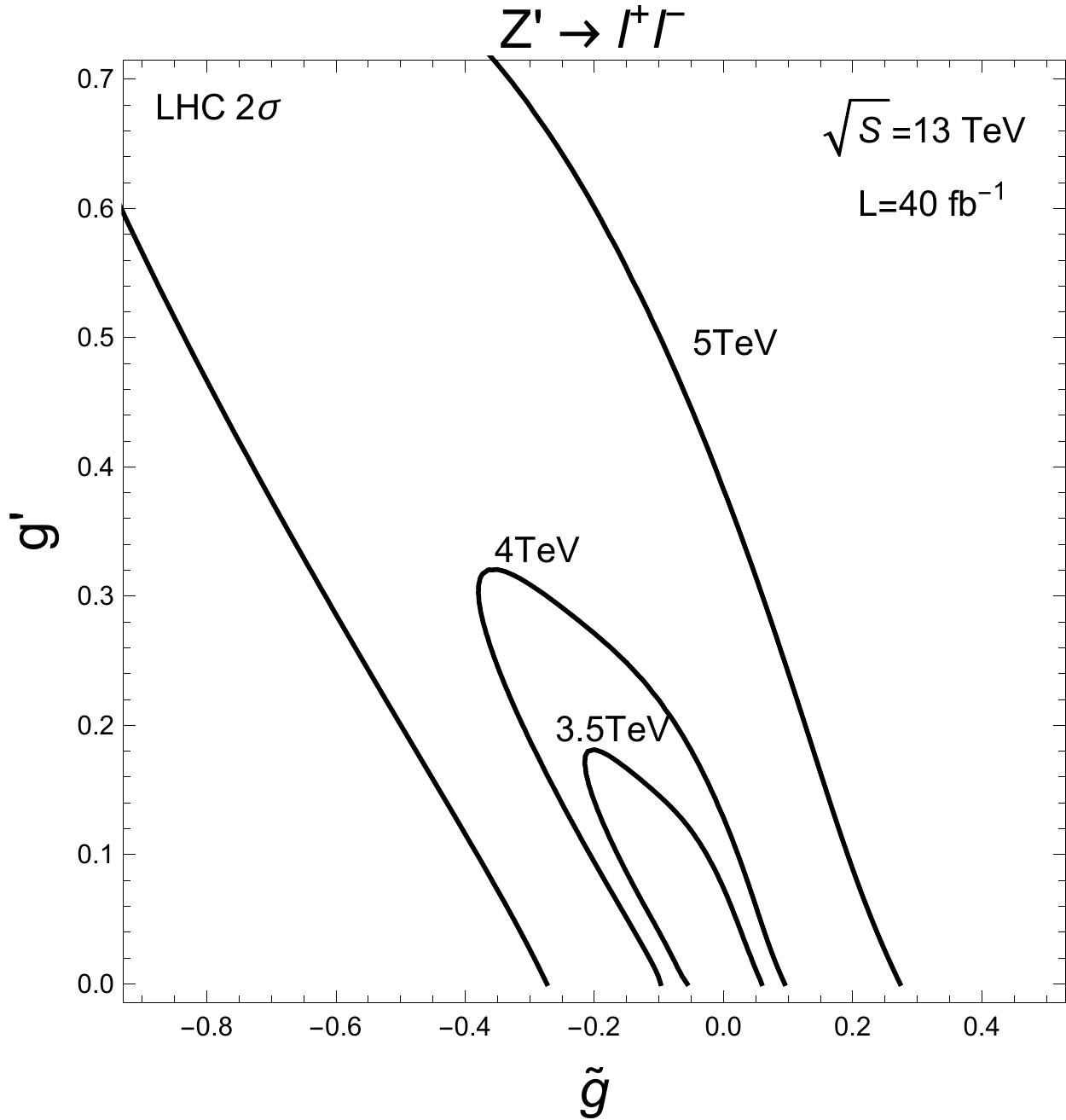}
\caption{Significance analysis for the di-lepton $(l = e, \mu)$ channel at the LHC for different $Z'$ masses and $\mathcal L = 40$ fb$^{-1}$.}
\label{exclusion}
\end{figure}

Before coming to that, we summarise here the main production mechanisms of heavy neutrinos, giving  an estimate of their expected cross section. Three are the main mediators for heavy neutrino pair production: the light SM-like Higgs, $H_1$, the heavy extra Higgs, $H_2$, and the heavy extra gauge boson $Z'$. The three contributing channels are:
\begin{itemize}
\item $gg\rightarrow H_1, H_2 \rightarrow \nu_h \nu_h$
\item $q\bar q \rightarrow Z, Z' \rightarrow \nu_h \nu_h$
\end{itemize}
As all cross sections depend either directly or indirectly on the mass and the coupling of the $Z'$-boson, in Fig.~\ref{exclusion} we display the 95\% C.L. bounds on the gauge coupling $g'$ and the mixing $\tilde g$ at fixed $Z'$-boson mass. We compute the 2$\sigma$ significance contour at the 13 TeV LHC, using the acceptance times efficiency factor given by the CMS analysis \cite{Khachatryan:2016zqb} and assuming a total luminosity $\mathcal L =40 \, \textrm{fb}^{-1}$. We combine the electron and muon channels. In addition, as the Higgs mediated cross section depends on the scalar mixing angle, $\alpha$, we take into account the exclusion limit on the $\alpha$ parameter as a function of the heavy Higgs mass. Such a limit has been extracted by using \texttt{HiggsBounds} \cite{arXiv:0811.4169,arXiv:1102.1898,arXiv:1301.2345,arXiv:1311.0055,arXiv:1507.06706} and \texttt{HiggsSignals} \cite{Bechtle:2013xfa} packages, see \cite{Accomando:2016sge}.

\begin{figure}[t]
\centering
\subfigure[]{\includegraphics[scale=0.57]{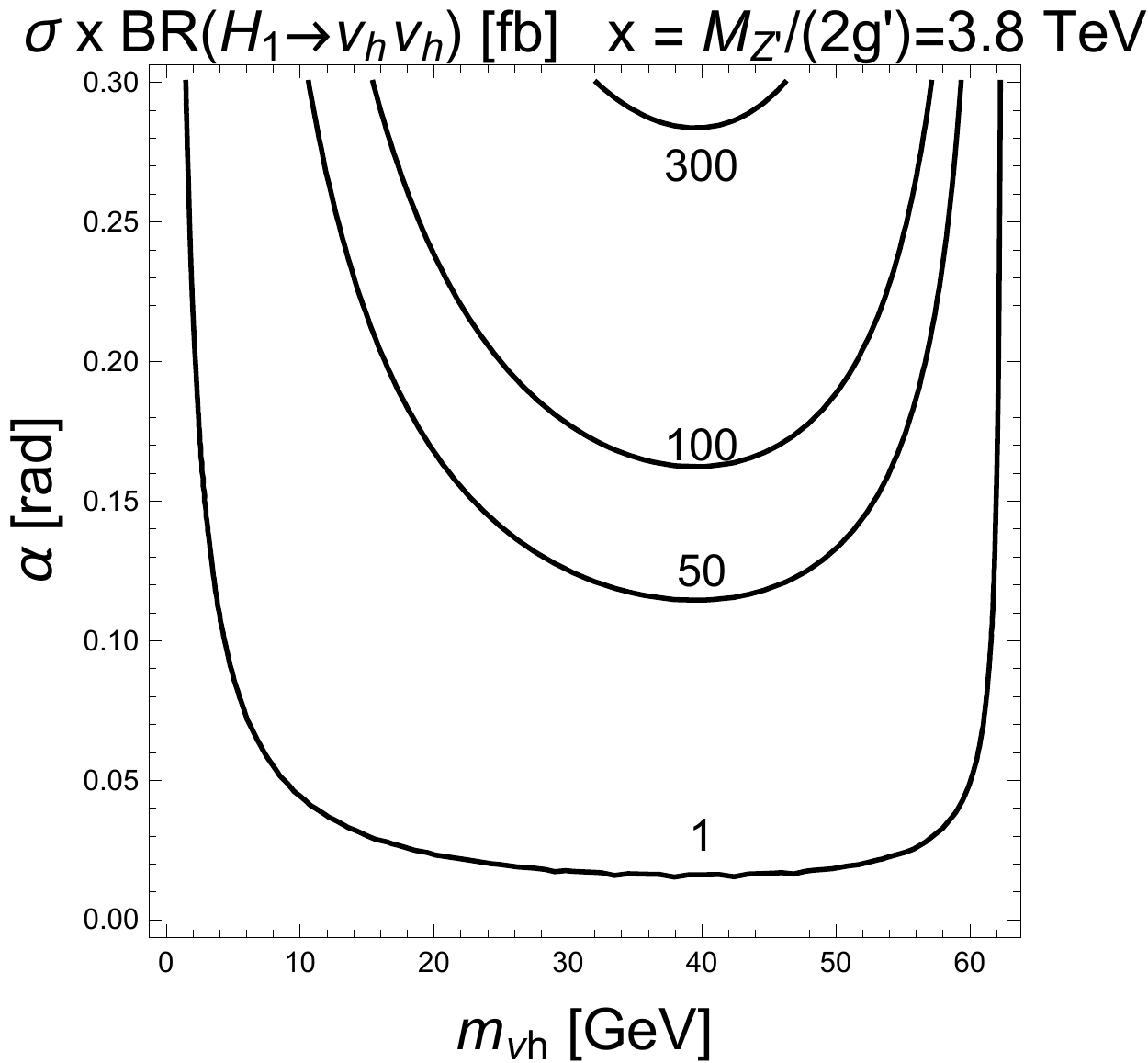}}
\subfigure[]{\includegraphics[scale=0.8]{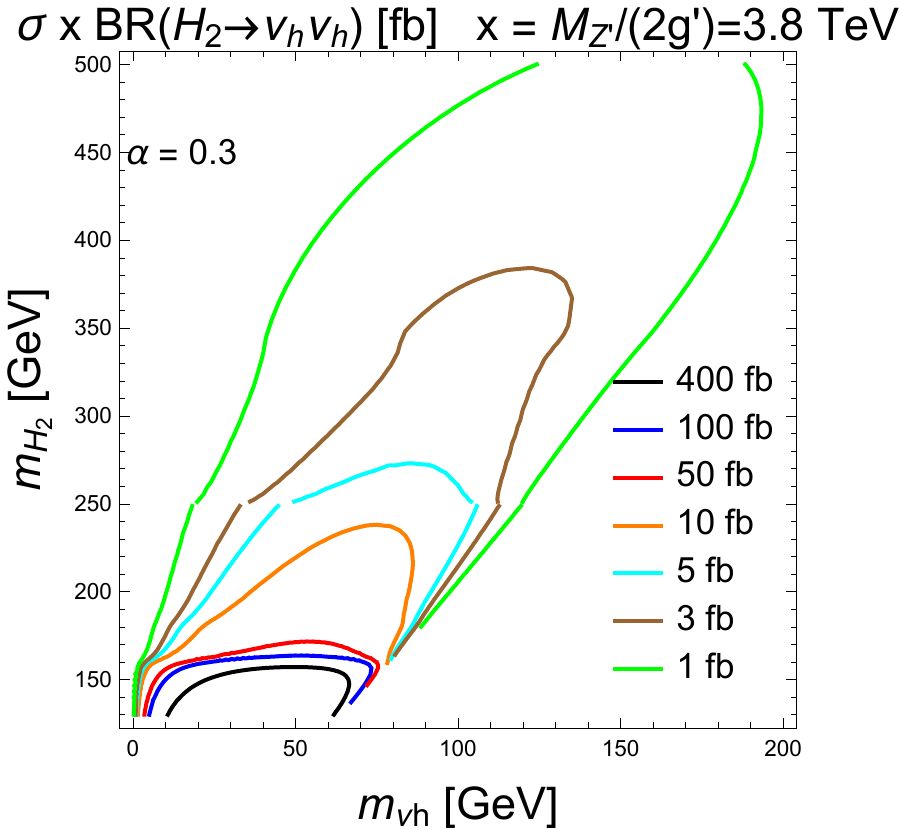}}
\caption{(a) Contour plots of the cross section times BR for the process $pp \rightarrow H_1 \rightarrow \nu_h \nu_h$ at the LHC with $\sqrt{S} = 13$ TeV in the $(m_{\nu_h}, \alpha)$ plane with $M_{Z'} = 5$ TeV and $g' \simeq 0.65$. (b) Contour plots of the cross section times BR for the process $pp \rightarrow H_2 \rightarrow \nu_h \nu_h$ at the LHC with $\sqrt{S} = 13$ TeV in the $(m_{\nu_h}, m_{H_2})$ plane for $M_{Z'} = 5$ TeV and $g' \simeq 0.65$. The parameter $x$ is the VEV of the extra scalar.}
\label{H12cross-section}
\end{figure}

When the allowed parameter space, we can compute the total cross sections for heavy neutrino production. In Fig.~\ref{H12cross-section}(a), we show the contour plots of the heavy neutrino pair production cross section via the sole light SM-like Higgs exchange in the plane $(m_{\nu_h}, \alpha )$.
We choose a value of the heavy Higgs VEV, $x = M_{Z'}/(2g') = 3.8$ TeV, which is allowed according to Fig.~\ref{exclusion}. The cross section heavily depends on $\alpha$, ranging from 1 to 300 fb with increasing $\alpha$. In this case of course the heavy neutrinos must be lighter than 60 GeV to be produced on mass shell. They are thus predominantly long-lived particles and give rise to a particular signature characterised by displaced vertices. In Fig.~\ref{H12cross-section}(b), we display the analogous cross section mediated by the sole heavy Higgs, $H_2$. In this case, we assume $\alpha = 0.3$ and we plot the cross section contours in the plane $(m_{\nu_h}, m_{H_2} )$. The value of the cross section can range between 1 and 400 fb, as before. It is maximal for a relatively light $H_2$ with mass $m_{H_2}\le 160$ GeV and consequently for relatively light neutrinos with $m_{\nu_h} \le 75$ GeV. Below this mass threshold, the $H_2$ branching fraction into heavy neutrinos is indeed substantial while above this threshold the $H_2\rightarrow W^+W^-$ channel opens up and becomes the dominant decay mode. For heavier neutrinos ($m_{\nu_h} \ge 75$ GeV) one can still have a rather sizeable production cross section via $H_2$ exchange. For $m_{\nu_h} \ge M_W$, when the heavy neutrinos are predominantly short-lived particles, the cross section can be of the order of a few fb, depending on the heavy Higgs mass (see brown and green contours).

\begin{figure}[t]
\centering
\subfigure[]{\includegraphics[scale=0.75]{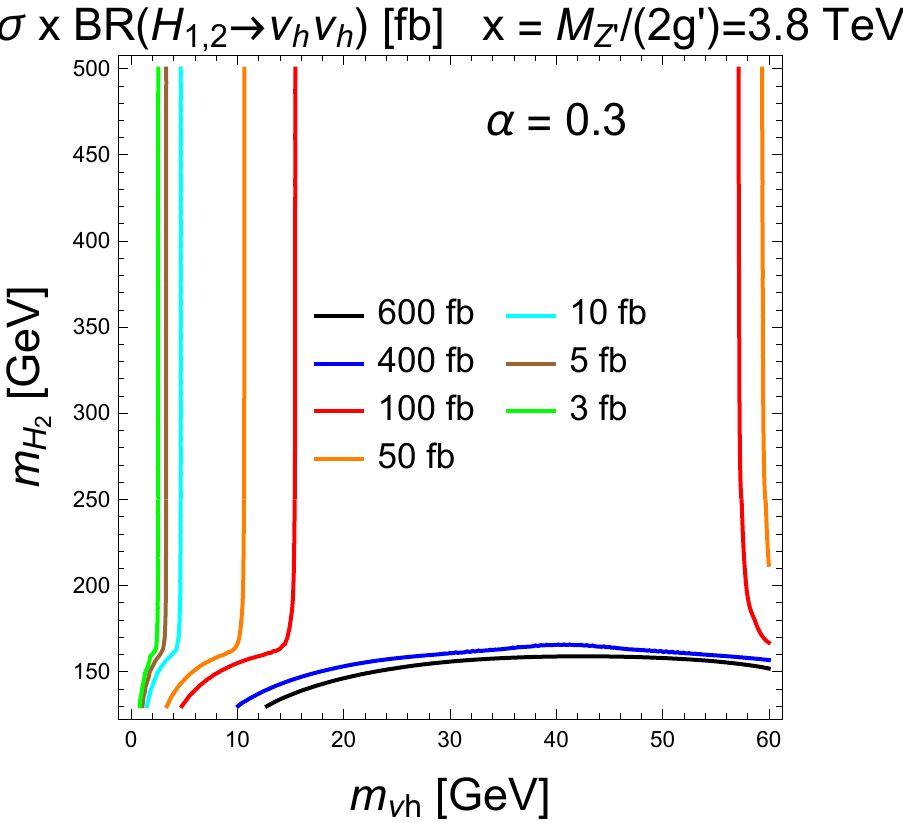}} \quad
\subfigure[]{\includegraphics[scale=0.75]{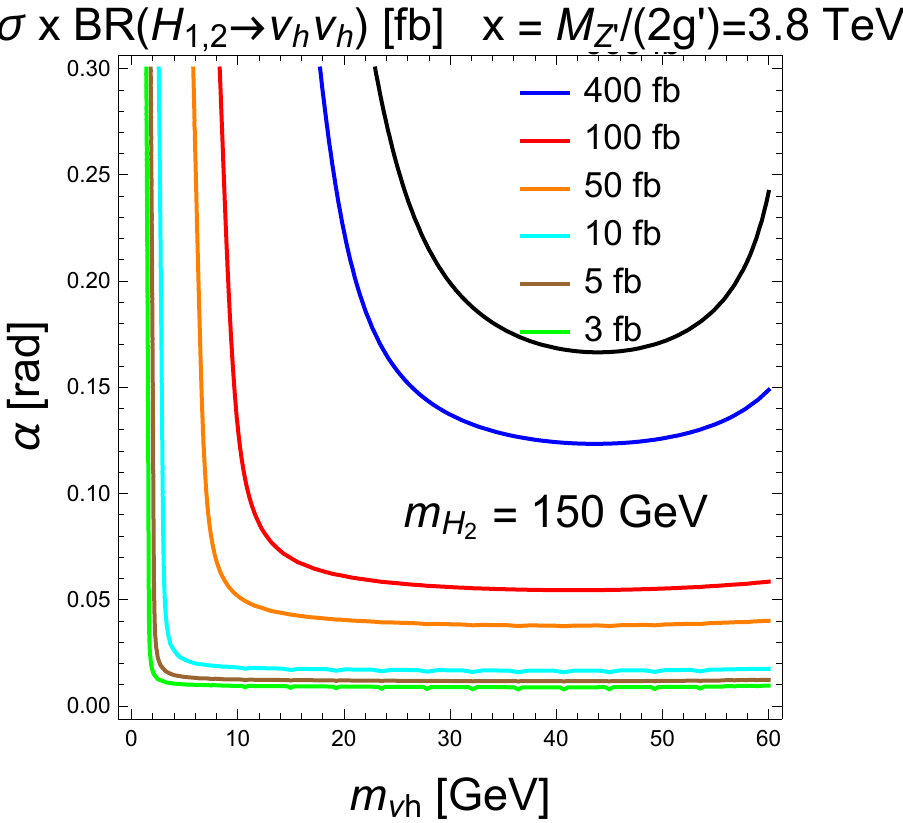}}
\caption{(a) Contour plots of the cross section times BR for the process $pp \rightarrow H_{1,2} \rightarrow \nu_h \nu_h$ at the LHC with $\sqrt{S} = 13$ TeV in the (a) $(m_{\nu_h}, m_{H_2})$ plane and (b)  $(m_{\nu_h}, \alpha)$ plane, with $M_{Z'} = 5$ TeV and $g' \simeq 0.65$. Notice that the cross sections for the channels induced by both $H_1$ and $H_2$ have been summed together. The interference is found to be negligible.}
\label{H12tot}
\end{figure}

As $H_1$ and $H_2$ can contribute to the same final state for $m_{\nu_h} \le 60$ GeV, a priori they could interfere each other. We have therefore computed the gluon-induced total cross section $gg \rightarrow H_1, H_2 \rightarrow \nu_h\nu_h$ including the interference between the two contributions. In Fig.~\ref{H12tot}, we show the total heavy neutrino production cross section via scalars in the planes $(m_{\nu_h}, m_{H_2} )$ (see plot (a) ) and $(m_{\nu_h}, \alpha )$ (see plot (b)). As one can see, the two contributions essentially sum up, with negligible interference. The total cross section reaches a maximum of about 700 fb  with a light additional Higgs, $m_{H_2}\le 150$ GeV, and a maximal $\alpha$ value. For $m_{\nu_h} \ge 60$ GeV, only $H_2$ contributes to the cross section whose value decreases with the heavy neutrino mass. For $m_{\nu_h} \ge 200$ GeV, the cross section drops to fractions of a fb.

\begin{figure}[t]
\centering
\subfigure[]{\includegraphics[scale=0.5]{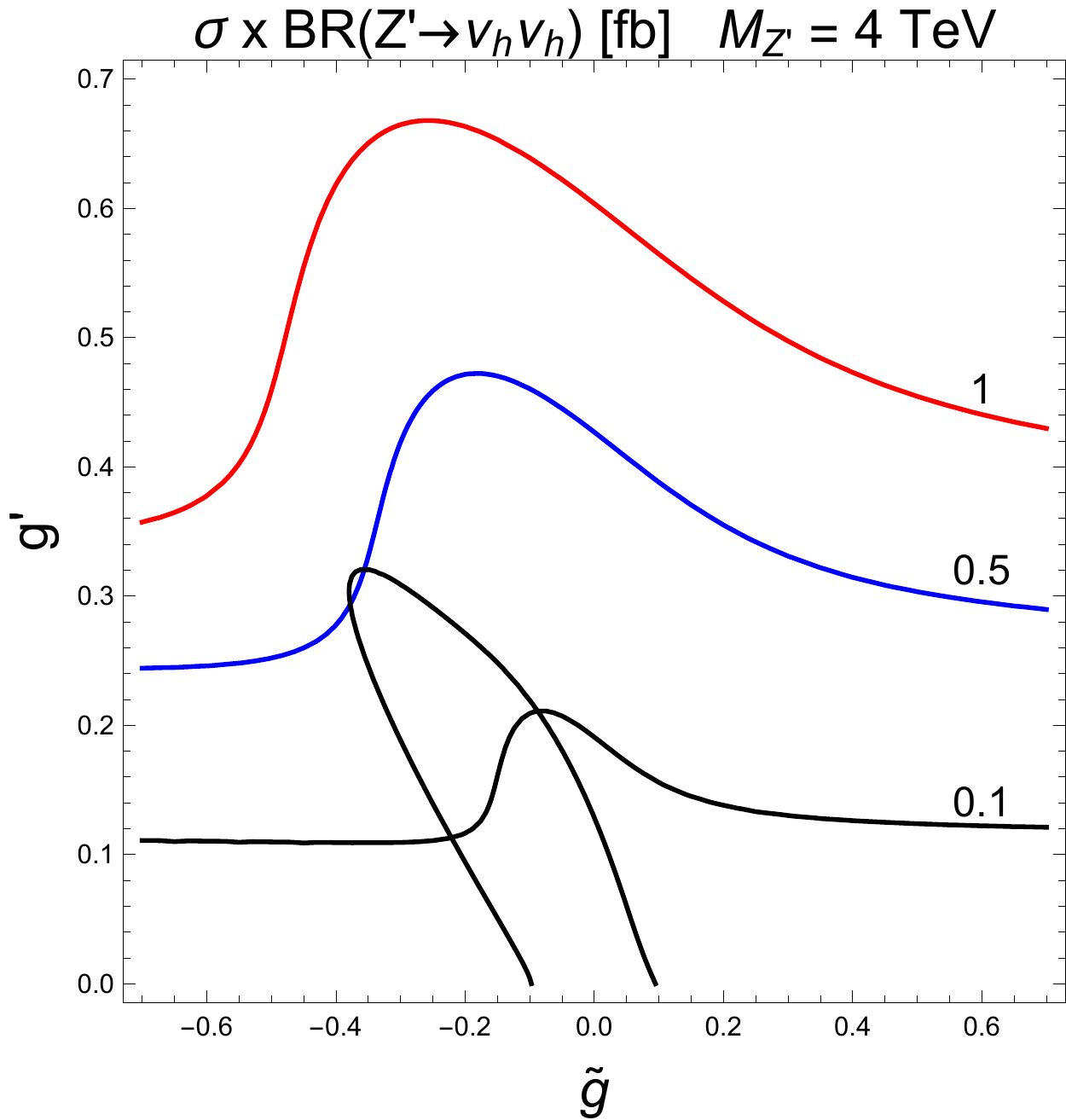}} \qquad
\subfigure[]{\includegraphics[scale=0.5]{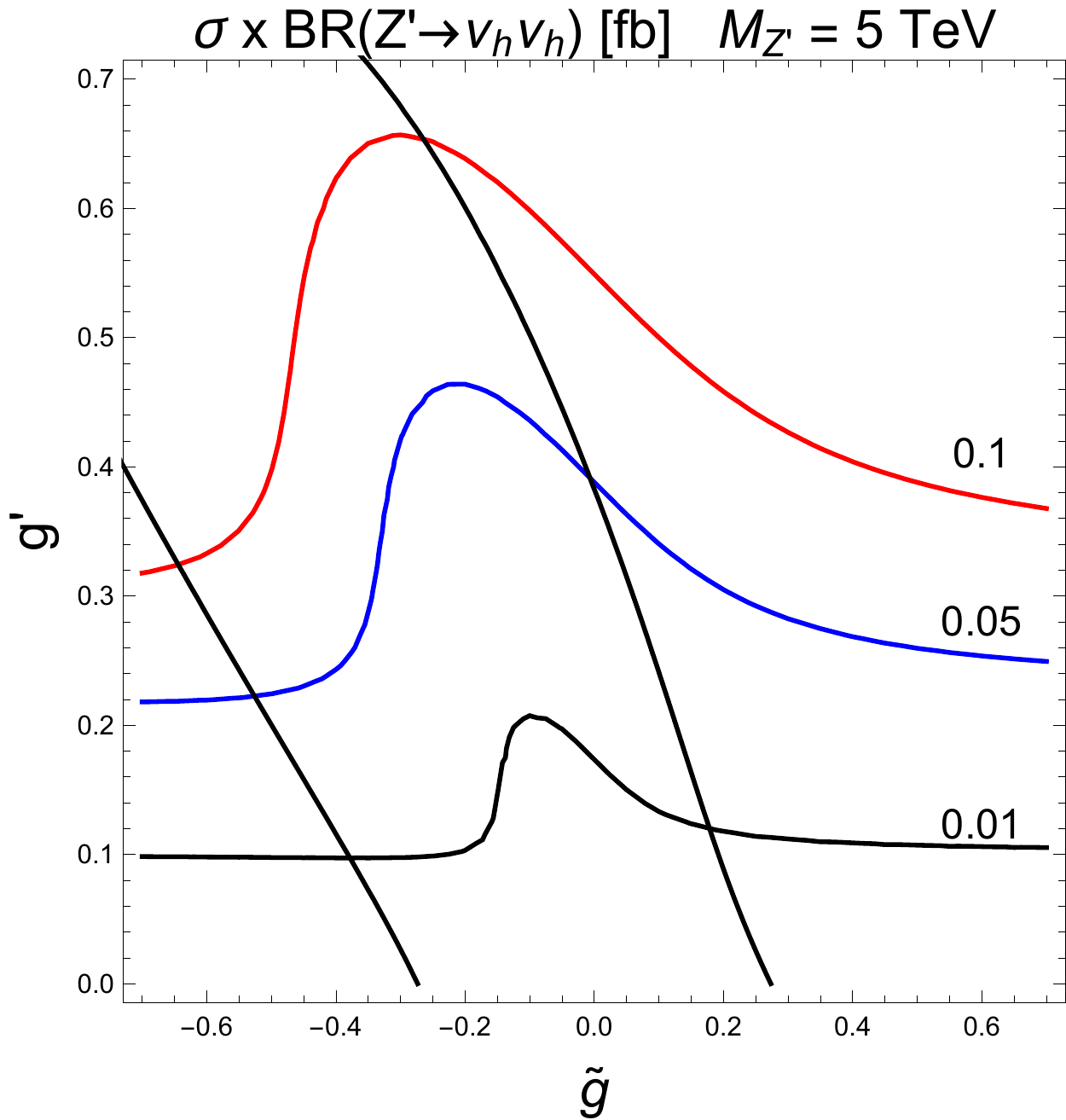}}
\caption{Contour plots of the cross section times BR for the process $pp \rightarrow Z' \rightarrow \nu_h \nu_h$ at the LHC with $\sqrt{S} = 13$ TeV in the $(\tilde g, g')$ plane. The limit $m_{\nu_h} \ll M_{Z'}$ has been considered.}
\label{Z'cross-section}
\end{figure}

Finally, we consider the quark-antiquark induced channel mediated by the neutral gauge bosons, $q\bar q \rightarrow Z, Z' \rightarrow \nu_h\nu_h$. As the $Z$-boson contribution is small, we neglect it.
In Fig.~\ref{Z'cross-section}, we show the contour plots of the heavy neutrino production cross section via the $Z'$-boson exchange in the plane $(\tilde g , g')$ for two values of the $Z'$ mass. The black contour with no label represents the 95\% C.L. exclusion limit at fixed $M_{Z'}$ from Fig.~\ref{exclusion}. Compared to the cross section mediated by the scalars, the gauge mediated cross section is generally much smaller. In this case, one cannot exceed a fraction of a femtobarn, independently of the heavy neutrino mass. Only in the high mass region of the neutrino spectrum, $m_{\nu_h} \gtrsim 200$ GeV, this channel can in principle be competitive with the scalar mediated one and for $m_{\nu_h} \gtrsim 300$ GeV becomes the main production mode.

In summary, we can identify different regions of the heavy neutrino and heavy scalar spectra where different pair production mechanisms dominate. In the regime where heavy neutrinos are long-lived, namely $m_{\nu_h} < M_W$, we delineate the following categories:
\begin{itemize}
\item $m_{\nu_h} < m_{H_1}/2$ and $m_{H_2} > 2 m_{W}$ \\
The heavy neutrinos are relatively light, $m_{\nu h}\le 60$ GeV and are produced from the decay of the SM-like Higgs. The heavier scalar predominantly decays into EW gauge bosons so that the heavy neutrino production cross section from $H_2$ is small compared to the one of the corresponding SM-like Higgs mediated channel. In this low mass region of the spectrum, the heavy neutrinos are predominantly long-lived and give rise to displaced vertices in the detector. This case has been studied in \cite{Accomando:2016rpc}.
\item $m_{\nu_h} > m_{H_1}/2$ and $2 m_{\nu_h} < m_{H_2} < 2 m_{W}$ \\
The heavy neutrinos are produced from the decay of the heavy Higgs with a sizeable cross section. For the maximal scalar mixing, $\alpha = 0.3$, the cross section can reach 400 fb. This region of the parameter space has been investigated in \cite{Accomando:2016rpc}. 
\item $m_{\nu_h} < m_{H_1}/2$ and $2 m_{\nu_h} < m_{H_2} < 2 m_{W}$ \\
The two scalars $H_1$ and $H_2$ both contribute to the heavy neutrino pair production. The interference between the two contributions is negligible, thus in first approximation the two cross sections simply sum up giving rise to a sizeable total cross section that for the maximal scalar mixing, $\alpha = 0.3$ approaches 700 fb.
 \item The two scalars are almost decoupled, $\alpha \simeq 0$ \\
Both the heavy scalar production cross section and the heavy neutrino decay mode of $H_2$ are suppressed so that the heavy neutrinos cannot be produced through the scalars. The $Z'$ production mode remains the only accessible channel despite its low cross section. 
\end{itemize}
In the short-lived heavy neutrino scenario we identify two main regions:
\begin{itemize}
\item $m_{\nu_h} < m_{H_2}/2$ and $m_{\nu_h} \lesssim 300$ GeV \\
The heavy Higgs is the main production mode for the heavy neutrinos with a cross section of a few fb.
\item $m_{\nu_h} > m_{H_2}/2$ or $m_{\nu_h} \gtrsim 300$ GeV \\
The heavy neutrino production from the $H_2$ is either kinematically forbidden or suppressed with respect to the $Z'$ channel. Despite its small cross section, the $Z'$ production mode through gauge interactions represents the main mechanism.  
\end{itemize}

In Ref.~\cite{Accomando:2016rpc}, we focused on the low to medium mass region of the heavy neutrino spectrum. There, the neutrino is very likely to be a long-lived particle giving rise to displaced vertices either in the inner tracker or in the muon chamber. The experimental techniques and the phenomenological analysis have been carried out in Ref.~\cite{Accomando:2016rpc}. 
In this paper, we address the medium to high mass region of the neutrino spectrum where the particle is dominantly short-lived. In order to analyse this case, we choose a final state consisting of three leptons, two jets and missing transverse energy.
 
\section{Short-lived heavy neutrinos: the $3l+2j+E_{T}^{\rm miss}$ signature}

In the region of parameter space where the heavy neutrinos are short-lived, namely for masses $m_{\nu_h} \gtrsim M_W$ GeV, standard reconstruction techniques can be successfully employed.
In this section we present a search for heavy neutrinos decaying into semi-leptonic final states characterised by three charged leptons (electrons and muons), two jets and missing energy at the LHC with a Centre-of-Mass energy of $\sqrt{S} = 13$ TeV. 
The heavy neutrinos can be pair produced by the heavy Higgs state or by the $Z'$ gauge boson, the latter being a signature property of abelian extensions of the SM.
The study of this signal is particularly useful as it allows the reconstruction of both the mass of the heavy neutrinos and the mass of the mediator at the same time. Indeed, fully-leptonic final states would be characterised by a smaller cross section and more missing energy as a lepton pair originating from a heavy neutrino is always accompanied by a light neutrino as required by the electric charge conservation. On the other hand, the final state with two leptons and four jets would suffer from a high SM background contamination. 
It is likely that if the mediator, either scalar or vector, is found at the LHC, it would first appear in other channels as DY for the $Z'$ and $\gamma\gamma$ for the heavy Higgs. 
Therefore the reconstruction of its mass in its heavy neutrino decay mode would represent an independent measurement suggesting the existence of extra SM neutral states participating in the generation of the light neutrino masses. \\
The two production channels are characterised by distinctive kinematical features, indeed while the $Z'$ mass limit is pushed beyond the 3 TeV region by searches in the di-lepton channel, the heavy Higgs mass is usually smaller than $\sim$ TeV to comply with the unitarity bounds. 
Therefore, if the heavy neutrino mass is below the TeV scale it is likely that is highly boosted in the former case, thus giving rise to decay products poorly separated in the angular direction. Fat jet and jet substructure techniques can be employed to search for these particles.\\
The main sources of background are represented by the $WZjj$ associated production, $t\bar t l \nu$ production and the $t \bar t$ pair production, the former representing the main contribution.
In the first case the three leptons come from the $W^\pm$ and $Z$ leptonic decays while in the $t\bar t l \nu$ case two jets and two leptons originate from the top-quark decay. In the last case the additional lepton is produced from the semi-leptonic decay of the $B$-meson. Notice that this lepton is not well isolated from the corresponding jet due to the large boost of the $b$ quark. This feature can be successfully exploited to suppressed an otherwise overwhelming background. \\
We discuss separately the two signals in the following sections.

\subsection{The heavy scalar mediator}

\begin{table}[t] \centering
\begin{tabular}{|c|ccc|} \hline
BP & $m_{H_2}$ (GeV) & $m_{\nu_h}$ (GeV) & $\sigma(\nu_h \nu_h)$ (fb) \\ \hline
BP1 & 250 & 100 & 6.50  \\
BP2 & 350 & 120 & 3.47  \\
BP3 & 450 & 180 & 1.32   \\
\hline
\end{tabular}
\caption{Benchmark points for the heavy neutrino pair production via the heavy scalar. The first and second columns give the heavy Higgs and the heavy neutrino masses, respectively. The third column displays the total cross section.
\label{BPs}}
\end{table}

We consider the three benchmark points in Tab.\ref{BPs} at a luminosity of $\mathcal L = 100 \, \textrm{fb}^{-1}$. The other parameters are chosen as follows $\alpha = 0.3$, $M_{Z'} = 5$ TeV, $g' = 0.65$ and comply with the exclusion bounds from Fig.\ref{exclusion} and from Higgs searches. 
The simulation has been performed with \texttt{CalcHEP} \cite{Belyaev:2012qa} using the $U(1)'$ model file \cite{Basso:2010jm, Basso:2011na} accessible on the High Energy Physics Model Data-Base (HEPMDB) \cite{hepmdb}.
The event samples generated by \texttt{CalcHEP} are then rescaled by the corresponding cross sections and the given luminosity. 
For the first benchmark, BP1, the heavy neutrino pair production cross section is found to be $\sigma_{pp \rightarrow H_2 \rightarrow \nu_h \nu_h} = 6.5$ fb where the heavy Higgs $H_2$ is produced through the gluon fusion process. With increasing $m_{H_2}$, the cross section drops down as displayed in Tab.~\ref{BPs} for the benchmark BP2. The two heavy neutrinos predominantly decay into leptons and jets through the $W^\pm$ and $Z$ gauge bosons by the following decay channels
\begin{itemize}
\item $\nu_h \rightarrow l^\mp W^\pm \rightarrow l^\mp l^{'\pm} \nu_{l'}$
\item $\nu_h \rightarrow l^\mp W^\pm \rightarrow l^\mp \bar q  q' $
\item $\nu_h \rightarrow \nu_{l'} Z \rightarrow \nu_{l'} l^+ l^- $
\item $\nu_h \rightarrow \nu_{l'} Z \rightarrow \nu_{l'} \bar q  q'  $
\item $\nu_h \rightarrow \nu_{l'} Z \rightarrow \nu_{l'} \nu_l \nu_l$
\end{itemize}
where $q$ can be one of the five light quarks and $l = e,\mu,\tau$ with $\tau$ decaying into $e,\mu$ and hadrons. \\

In order to perform our analysis, we first employ generic detector acceptance requirements to identify and reconstruct the three leptons and the two jets in the final state. As there are different trigger thresholds for different lepton flavour combinations, we need to divide the final state into four categories: $3\mu$, $2\mu e$, $2e \mu$, $3e$ ($+2j+E_{T}^{\rm miss}$). To reduce the turn on effect, we impose cuts a couple of GeV above the corresponding threshold. The transverse momentum cut in each category is summarised below:

\begin{itemize}
\item $3\mu + 2j + E_{T}^{\rm miss}$: $p^T_{\mu_{1}} > 14$ GeV, $p^T_{\mu_{2}} > 12$ GeV, $p^T_{\mu_{3}} > 7$ GeV
\item $2\mu + e + 2j + E_{T}^{\rm miss}$: $p^T_{l} > 11$ GeV
\item $\mu + 2e + 2j + E_{T}^{\rm miss}$: $p^T_{\mu} > 10$ GeV, $p^T_{e} > 14$ GeV
\item $3e + 2j + E_{T}^{\rm miss}$: $p^T_{e_{1, 2}} > 28$ GeV, $p^T_{e_{3}} > 5$ GeV
\end{itemize}
 
In addition, we impose a cut on the pseudo-rapidity of each lepton equal to $|\eta_l| < 2.5$.
The two jets are characterised, instead, by $p^T_{j_{1,2}} > 30$ GeV and  $|\eta_j| < 3$. 
Concerning the isolation of the different objects we require $\Delta R_{ll} > 0.3$ between any two leptons, $\Delta R_{jj} > 0.4$ for the two jets and $\Delta R_{lj} > 0.4$
between any lepton and jet. The corresponding efficiencies are shown in Tabs.~\ref{fig.H2BP1.1} - \ref{fig.H2BP2.2} for the two chosen benchmark points.

\begin{figure}[t]
\centering
\subfigure[]{\includegraphics[scale=0.3]{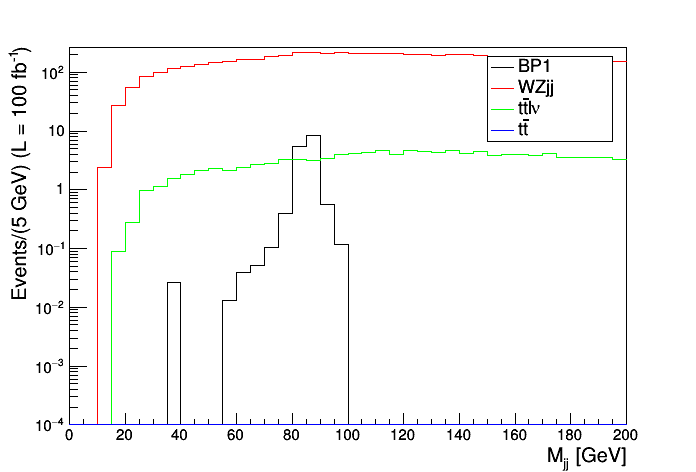}} 
\subfigure[]{\includegraphics[scale=0.3]{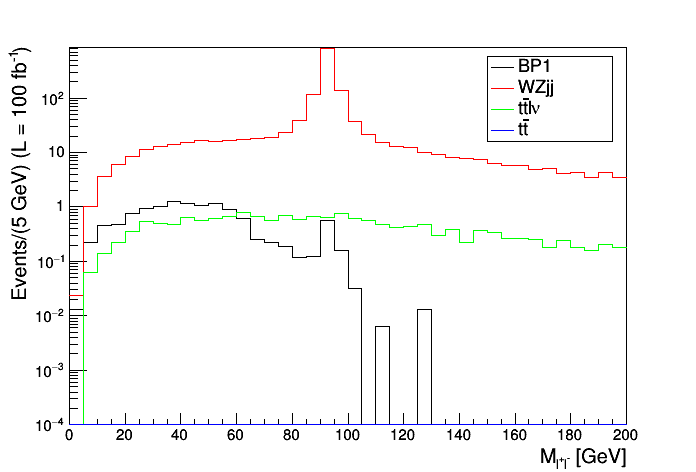}} 
\caption{(a) Distribution in the di-jet invariant mass for the benchmark point BP1. Acceptance cuts are applied. (b) Distribution in the di-lepton invariant mass for the benchmark point BP1. Acceptance cuts and $|M_{jj} - M_W| < 20$ GeV are applied. A luminosity $\mathcal L = 100$ fb$^{-1}$ is assumed.
\label{fig.BP1dist}}
\end{figure}

The smallest efficiency of the selection conditions of the signal events is related to the the $p^T$ cuts due to the relative small mass of the heavy neutrinos that undergo a three-body decay into leptons and jets. For instance, in the BP1 case, the three-electron category is particularly affected by the trigger thresholds imposed, with an efficiency of the order of 6\%. The three muon category has the best efficiency of about 30\%, owing to the lowest asymmetric $p^T$ cuts applied. Within all four categories, the decay objects are adequately isolated and therefore the angular separation cuts do not considerably affect the signal. As for the SM background, the main contribution comes from three processes: $WZjj$, $t\bar t$, and $t\bar t l\nu_l$. The SM background is dominant compared to the heavy neutrino pair production signal, as shown in Tabs.~\ref{fig.H2BP1.1} - \ref{fig.H2BP2.2}, even after acceptance cuts which significantly reduce the background. 
In particular, the $\Delta R_{lj}$ cut is quite effective in the suppression of the $t \bar t$ background as in this process the additional lepton coming from the $B$-meson decay is produced very close to the jet from which it originates. Yet, after the acceptance cuts, the SM background is typically three orders of magnitude larger than the signal. The implementation of other kinematical cuts is thus needed to make the search for heavy neutrinos viable.

\begin{figure}[t]
\centering
\subfigure[]{\includegraphics[scale=0.27]{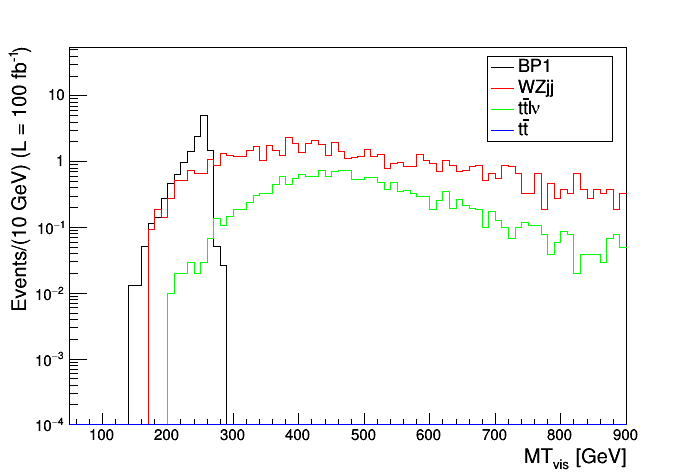}} \qquad
\subfigure[]{\includegraphics[scale=0.27]{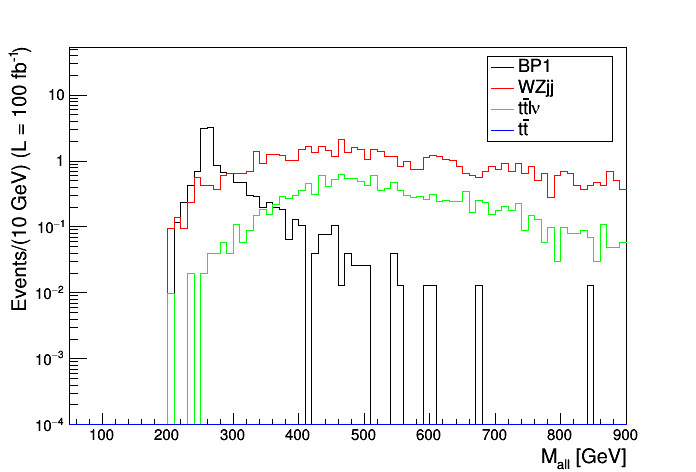}}
\caption{(a) Distribution in the visible transverse mass of the final state. (b) Distribution in the total invariant mass of the final state obtained via a light neutrino reconstruction algorithm, as explained in the text. Acceptance cuts plus $|M_{jj} - M_W| < 20$ GeV and $|M_{l^+l^-} - M_Z| > 20$ GeV have been applied (see text). A luminosity $\mathcal L = 100$ fb$^{-1}$ is assumed.
\label{fig.BP1masses}}
\end{figure}

Since the two jets coming from the three-body decay of one of the two heavy neutrinos are always produced by the decay of a $W^\pm$-boson, one can require their invariant mass to be close to $M_W$, namely $|M_{jj} - M_W| < 20$ GeV. 
This cut effectively reduces the background but it is not sufficient to distinguish the signal and the events from the $WZjj$ process still represent the main source of SM contamination.
In order to improve the signal significance we require $|M_{l^+l^-} - M_Z| > 20$ GeV where $M_{l^+l^-}$ is the invariant mass of all the lepton pairs with opposite charge and same flavour, thus potentially identifying and removing the SM $Z$-gauge boson. \\
In order to visualise the strong impact of the reducible SM background and the effect of these kinematical cuts, in Fig.~\ref{fig.BP1dist}(a) we show the distribution in the di-jet invariant mass. Here we sum up the number of events from all four categories, after the corresponding acceptance cuts defined above are applied. In Fig.~\ref{fig.BP1dist}(b) we display the invariant mass of all the same-flavour and opposite-sign lepton pairs. This time, the additional cut $|M_{jj} - M_W| < 20$ GeV is superimposed on the acceptance constraints. The green line represents the $t\bar t l\nu$ while the red line is the $WZjj$ background. The blue line, the $t \bar t$ background, has been reduced to zero due to the $\Delta R_{lj}>0.4$ requirement. After implementing the suggested kinematical cuts on the di-jet and di-lepton invariant masses, the SM background drops down by two orders of magnitude and more. This results in the background being roughly three times bigger than the signal for the $3\mu$-category (see Tabs.~\ref{fig.H2BP1.1} - \ref{fig.H2BP2.2}).

\begin{figure}[t]
\centering
\includegraphics[scale=0.38]{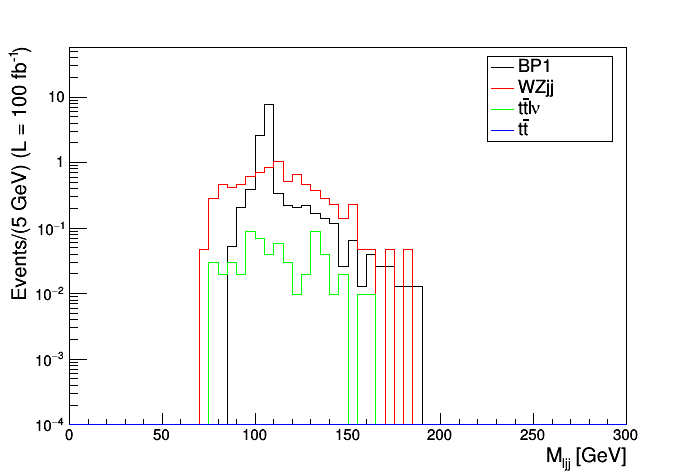}
\caption{Distribution in the invariant mass of the two jets and the closest lepton, $M_{ljj}$, for the benchmark point BP1. Acceptance cuts plus $|M_{jj} - M_W| < 20$ GeV, $|M_{l^+l^-} - M_Z| > 20$ GeV and $|M^T_{vis} - m_{H_{2}}| < 50$ GeV have been applied (see text). A luminosity $\mathcal L = 100$ fb$^{-1}$ is assumed.
\label{fig.BP1dist2}}
\end{figure}

In Fig.~\ref{fig.BP1masses}, we show two invariant mass distributions after the cuts discussed above. These observables can be used to extract the heavy Higgs mass. In particular, we show the transverse mass of all the visible particles $M^T_\textrm{vis}$ and the invariant mass of all particles $M_{all}$ including the light neutrino whose longitudinal momentum can be reconstructed through the $W$ mass. Here, we sum up all events from the four above-mentioned categories. We then focus on the major aim of our paper that is reconstructing the heavy neutrino mass.
From the shape of the $M^T_\textrm{vis}$ distribution, one realises that a cut on this variable can help enhancing the signal over the SM background. We then impose that $|M^T_{vis} - m_{H_2}| < 50$ GeV. With this constraint, the reducible SM background decreases by a factor of ten while the signal remains almost unaffected. The consequence is twofold. The significance increases by a factor of four, if one combines the four categories shown in Tabs.~\ref{fig.H2BP1.1} - \ref{fig.H2BP1.2}, going from $S/\sqrt{B} = 1.4$ to $S/\sqrt{B} = 4.3$. In addition, using the invariant mass of the two jets and the closest lepton $M_{ljj}$, the reconstruction of the heavy neutrino mass is much clearer as shown in Fig.~\ref{fig.BP1dist2} where all four channels are combined. Analogous results are obtained when applying the cut $|M_{all} - m_{H_2}| < 50$ GeV with a final significance of $S/\sqrt{B} = 5.1$.

In the benchmark point BP2 of Tab.~\ref{BPs}, the heavy Higgs and the heavy neutrino masses  increase. The neutrino pair production cross section goes down by roughly a factor of two, compared to the BP1 case, owing to the higher energy scale in the PDF's and to the opening of the $t\bar t$ decay channel for the heavy Higgs. In contrast, the trigger thresholds efficiency increases, because the heavy neutrino more massive. This feature is shown in Tabs. \ref{fig.H2BP2.1} - \ref{fig.H2BP2.2}. It is particularly striking for the $eee$ category, characterised by the highest $p^T$ cuts, where the efficiency raises from 6\% (for BP1) to 30\% (for BP2) just with a 20 GeV increase in the heavy neutrino mass. Altogether, the significance of the four combined channels is $S/\sqrt{B} = 1.5$ if the $M^T_\textrm{vis}$ cut is used or $S/\sqrt{B} = 1.7$ if the $M_{all}$ cut is employed instead, for a luminosity  
$\mathcal L = 100$ fb$^{-1}$. If we increase the heavy Higgs and the heavy neutrino mass even further, the sensitivity goes down, of course. In order to estimate the maximum $m_{\nu h}$ that one could explore in this channel, we have evaluated the BP3 reference point in Tab.~\ref{BPs}. Here we get three events at $\mathcal L = 100$ fb$^{-1}$ and a significance $S/\sqrt{B} = 0.5$. Projecting to the High Luminosity Large Hadron Collider (HL-LHC) stage with $\mathcal L = 3$ ab$^{-1}$, one reaches a significance $S/\sqrt{B} = 2.7$. The analysis in this channel is thus appropriate for the search of a heavy neutrino in the mass range 100 GeV $\le m_{\nu h} \le $ 180 GeV.

In the next section, we study the heavy neutrino pair production mediated by the heavy $Z'$-boson. This channel becomes complementary when the heavy neutrino is rather massive, $m_{\nu h} > $ 180 GeV, or in the decoupling limit when $\alpha$ = 0.
 
\begin{sidewaystable}[t]
\begin{tabular}{|c||c|c||c|c||c|c||c|c||c|} \hline
				& BP1($eee$)	& Eff. \%	& $WZjj$	& Eff. \%	& $t \bar t l \nu$	& Eff. \%	& $t \bar t$	& Eff. \% 	& $S/\sqrt{B}$			\\ \hline \hline
No cuts.			 & 36.3692	 & 100			 & 18271.1	 & 100			 & 60.0229	 & 100			 & 7064.84	 & 100 		& 0.228219 \\
$\eta$			 & 26.7929	& 73.67			 & 9292.88	& 50.86			 & 42.1034	& 70.15			 & 5383.75	& 76.2		& 0.220844 \\
$p^{T}$			 & 1.74115	& 6.499			 & 3646.03	& 39.23			 & 25.8532	& 61.4			 & 1194.66	& 22.19 		& 0.0249589 \\
$\Delta R$ 			 & 1.4293	& 82.09			 & 3174.62	& 87.07			 & 21.8961	& 84.69			 & 0	& 0 					& 0.0252805 \\
$|M_{jj} - M_W| < 20$ GeV			 & 1.4293	& 100			 & 379.156	& 11.94			 & 3.19114	& 14.57			 & 0	& 100 	& 0.0730963 \\
$|M_{l^+l^-} - M_Z| > 20$ GeV			 & 1.0135	& 70.91			 & 11.0076	& 2.903			 & 1.80668	& 56.62			 & 0	& 100 	& 0.283125 \\
\hline \hline
$|M^T_{vis} - m_{H_{2}}| < 50$ GeV			 & 1.00051	& 98.72			 & 0.60635	& 5.508			 & 0.0490945	& 2.717			 & 0	& 100  & 1.23582 \\
$|M_{all} - m_{H_{2}}| < 50$ GeV			 & 0.844587	& 83.33			 & 0.233212	& 2.119			 & 0.0098189	& 0.5435			 & 0	& 100 & 1.71322 \\ 
\hline
\end{tabular}
\\
\bigskip \bigskip \bigskip \\
\begin{tabular}{|c||c|c||c|c||c|c||c|c||c|} \hline
				& BP1($ee\mu$)	& Eff. \%	& $WZjj$	& Eff. \%	& $t \bar t l \nu$	& Eff. \%	& $t \bar t$	& Eff. \% 	& $S/\sqrt{B}$			\\ \hline \hline
No cuts.			 & 38.89	 & 100			 & 19574.3	 & 100			 & 185.155	 & 100			 & 21486.4	 & 100	& 0.191491 \\
$\eta$			 & 28.625	& 73.61			 & 9957.39	& 50.87			 & 130.277	& 70.36			 & 16520.9	& 76.89	& 0.175483 \\
$p^{T}$			 & 3.70319	& 12.94			 & 3652.42	& 36.68			 & 76.7151	& 58.89			 & 3164.68	& 19.16	& 0.0446012 \\
$\Delta R$			 & 3.14446	& 84.91			 & 3177.7	& 87			 & 65.0109	& 84.74			 & 0	& 0	& 0.0552195 \\
$|M_{jj} - M_W| < 20$ GeV			 & 3.13147	& 99.59			 & 377.85	& 11.89			 & 8.31661	& 12.79			 & 0	& 100	& 0.159353 \\
$|M_{l^+l^-} - M_Z| > 20$ GeV			 & 2.80663	& 89.63			 & 21.3155	& 5.641			 & 6.79468	& 81.7			 & 0	& 100	& 0.529362 \\
\hline \hline
$|M^T_{vis} - m_{H_{2}}| < 50$ GeV			 & 2.70268	& 96.3			 & 2.70526	& 12.69			 & 0.147283	& 2.168			 & 0	& 100	& 1.60022 \\
$|M_{all} - m_{H_{2}}| < 50$ GeV			 & 2.2479	& 80.09			 & 1.35263	& 6.346			 & 0.0589134	& 0.8671			 & 0	& 100	& 1.89204 \\
\hline
\end{tabular}
\caption{Benchmark point BP1. Luminosity $\mathcal L = 100$ fb$^{-1}$.
Universal acceptance cuts = $|\eta_l| < 2.5$, $|\eta_j| < 3$, $p^T_{j_{1,2}} > 30$ GeV,  $\Delta R_{jj} > 0.4$ + $\Delta R_{lj} > 0.4$ + $\Delta R_{ll} > 0.3$. Transverse momentum cuts for the $eee$ category: 
$p^T_{e_{1, 2}} > 28$ GeV and $p^T_{e_{3}} > 5$ GeV. Transverse momentum cuts for the $ee\mu$ category: $p^T_{e_{1, 2}} > 14$ GeV and $p^T_{\mu} > 10$ GeV. Notice that here and in all the following tables, 0 events means the estimated value is less than $10^{-4}$. }
\label{fig.H2BP1.1}
\end{sidewaystable}

\begin{sidewaystable}[t]
\begin{tabular}{|c||c|c||c|c||c|c||c|c||c|} \hline
				& BP1($e\mu\mu$)	& Eff. \%	& $WZjj$	& Eff. \%	& $t \bar t l \nu$	& Eff. \%	& $t \bar t$	& Eff. \% 	& $S/\sqrt{B}$			\\ \hline \hline
				No cuts.			 & 38.0454	 & 100			 & 19613.8	 & 100			 & 187.639	 & 100			 & 21109.9	 & 100	& 0.188096 \\
$\eta$			 & 27.7155	& 72.85			 & 10031.1	& 51.14			 & 132.82	& 70.78			 & 16298.2	& 77.21	& 0.170377 \\
$p^{T}$			 & 5.17147	& 18.66			 & 3776.77	& 37.65			 & 81.6147	& 61.45			 & 3599.54	& 22.09	& 0.0598832 \\
$\Delta R$			 & 4.08001	& 78.89			 & 3287.72	& 87.05			 & 69.8418	& 85.58			 & 0	& 0	& 0.0704123 \\
$|M_{jj} - M_W| < 20$ GeV			 & 4.05402	& 99.36			 & 394.687	& 12			 & 9.82872	& 14.07			 & 0	& 100	& 0.201566 \\
$|M_{l^+l^-} - M_Z| > 20$ GeV			 & 3.91109	& 96.47			 & 25.6999	& 6.511			 & 8.02204	& 81.62			 & 0	& 100	& 0.673506 \\
\hline \hline
$|M^T_{vis} - m_{H_{2}}|  < 50$ GeV			 & 3.70319	& 94.68			 & 2.70526	& 10.53			 & 0.314205	& 3.917			 & 0	& 100	& 2.13114 \\
$|M_{all} - m_{H_{2}}|  < 50$ GeV			 & 2.92357	& 74.75			 & 1.25934	& 4.9			 & 0.117827	& 1.469			 & 0	& 100	& 2.49127 \\
\hline
\end{tabular}
\\
\bigskip \bigskip \bigskip \\
\begin{tabular}{|c||c|c||c|c||c|c||c|c||c|} \hline
				& BP1($\mu\mu\mu$)	& Eff. \%	& $WZjj$	& Eff. \%	& $t \bar t l \nu$	& Eff. \%	& $t \bar t$	& Eff. \% 	& $S/\sqrt{B}$			\\ \hline \hline
No cuts.			 & 35.6546	 & 100			 & 18102	 & 100			 & 64.9422	 & 100			 & 6901.4	 & 100	& 0.225192 \\
$\eta$			 & 26.4941	& 74.31			 & 9214.47	& 50.9			 & 45.9328	& 70.73			 & 5336.08	& 77.32	& 0.219293 \\
$p^{T}$			 & 7.74422	& 29.23			 & 3910.77	& 42.44			 & 32.4907	& 70.74			 & 1428.14	& 26.76	& 0.105666 \\
$\Delta R$			 & 6.5488	& 84.56			 & 3401.86	& 86.99			 & 27.7482	& 85.4			 & 0	& 0	& 0.111825 \\
$|M_{jj} - M_W| < 20$ GeV			 & 6.5488	& 100			 & 413.624	& 12.16			 & 3.85883	& 13.91			 & 0	& 100	& 0.320511 \\
$|M_{l^+l^-} - M_Z| > 20$ GeV			 & 5.43135	& 82.94			 & 15.252	& 3.687			 & 2.16016	& 55.98			 & 0	& 100	& 1.30161 \\
\hline\hline
$|M^T_{vis} - m_{H_{2}}| < 50$ GeV				 & 5.13249	& 94.5			 & 1.91234	& 12.54			 & 0.0785512	& 3.636			 & 0	& 100	& 3.63752 \\
$|M_{all} - m_{H_{2}}|  < 50$ GeV				 & 3.96306	& 72.97			 & 0.746277	& 4.893			 & 0.0392756	& 1.818			 & 0	& 100	& 4.4714 \\
\hline
\end{tabular}
\caption{Benchmark point BP1. Luminosity $\mathcal L = 100$ fb$^{-1}$.
Universal acceptance cuts = $|\eta_l| < 2.5$, $|\eta_j| < 3$, $p^T_{j_{1,2}} > 30$ GeV,  $\Delta R_{jj} > 0.4$ + $\Delta R_{lj} > 0.4$ + $\Delta R_{ll} > 0.3$. Transverse momentum cuts for the $e\mu\mu$ category: $p^T_{l_{1, 2, 3}} > 11$ GeV ($l = e, \mu$). Transverse momentum cuts for the $\mu\mu\mu$ category: $p^T_{\mu_{1}} > 14$ GeV,  $p^T_{\mu_{2}} > 12$ GeV and $p^T_{\mu_3} > 7$ GeV. }
\label{fig.H2BP1.2}
\end{sidewaystable}


\begin{sidewaystable}[t]
\begin{tabular}{|c||c|c||c|c||c|c||c|c||c|} \hline
				& BP2($eee$)	& Eff. \%	& $WZjj$	& Eff. \%	& $t \bar t l \nu$	& Eff. \%	& $t \bar t$	& Eff. \% 	& $S/\sqrt{B}$			\\ \hline \hline
No cuts.			 & 13.9652	 & 100			 & 18271.1	 & 100			 & 60.0229	 & 100			 & 7064.84	 & 100 		& 0.0876327 \\
$\eta$			 & 10.6385	& 76.18			 & 9292.88	& 50.86			 & 42.1034	& 70.15			 & 5383.75	& 76.2 		& 0.0876893 \\
$p^{T}$			 & 3.21582	& 30.23			 & 3646.03	& 39.23			 & 25.8532	& 61.4			 & 1194.66	& 22.19 		& 0.0460979 \\
$\Delta R$			 & 2.61978	& 81.47			 & 3174.62	& 87.07			 & 21.8961	& 84.69			 & 0	& 0 				& 0.0463369 \\
$|M_{jj} - M_W| < 20$ GeV			 & 2.57127	& 98.15			 & 379.156	& 11.94			 & 3.19114	& 14.57			 & 0	& 100 	& 0.131498 \\
$|M_{l^+l^-} - M_Z| > 20$ GeV			 & 0.873261	& 33.96			 & 11.0076	& 2.903			 & 1.80668	& 56.62			 & 0	& 100 	& 0.243948 \\
\hline \hline
$|M^T_{vis} - m_{H_{2}}| < 50$ GeV			 & 0.797024	& 91.27			 & 2.00562	& 18.22			 & 0.274929	& 15.22			 & 0	& 100 	& 0.527778 \\
$|M_{all} - m_{H_{2}}| < 50$ GeV			 & 0.741578	& 84.92			 & 1.16606	& 10.59			 & 0.147283	& 8.152			 & 0	& 100 	& 0.647095 \\
\hline
\end{tabular}
\\
\bigskip \bigskip \bigskip \\
\begin{tabular}{|c||c|c||c|c||c|c||c|c||c|} \hline
				& BP2($ee\mu$)	& Eff. \%	& $WZjj$	& Eff. \%	& $t \bar t l \nu$	& Eff. \%	& $t \bar t$	& Eff. \% 	& $S/\sqrt{B}$			\\ \hline \hline
No cuts.			 & 15.4415	 & 100			 & 19574.3	 & 100			 & 185.155	 & 100			 & 21486.4	 & 100	& 0.0760323 \\
$\eta$			 & 12.0316	& 77.92			 & 9957.39	& 50.87			 & 130.277	& 70.36			 & 16520.9	& 76.89	& 0.0737585 \\
$p^{T}$			 & 3.68017	& 30.59			 & 3652.42	& 36.68			 & 76.7151	& 58.89			 & 3164.68	& 19.16	& 0.0443239 \\
$\Delta R$			 & 3.22968	& 87.76			 & 3177.7	& 87			 & 65.0109	& 84.74			 & 0	& 0	& 0.056716 \\
$|M_{jj} - M_W| < 20$ GeV			 & 3.19503	& 98.93			 & 377.85	& 11.89			 & 8.31661	& 12.79			 & 0	& 100	& 0.162588 \\
$|M_{l^+l^-} - M_Z| > 20$ GeV			 & 2.28711	& 71.58			 & 21.3155	& 5.641			 & 6.79468	& 81.7			 & 0	& 100	& 0.431375 \\
\hline \hline
$|M^T_{vis} - m_{H_{2}}| < 50$ GeV			 & 2.07226	& 90.61			 & 4.10453	& 19.26			 & 1.4532	& 21.39			 & 0	& 100	& 0.879015 \\
$|M_{all} - m_{H_{2}}| < 50$ GeV			 & 1.8782	& 82.12			 & 2.51869	& 11.82			 & 0.638228	& 9.393			 & 0	& 100	& 1.05709 \\
\hline
\end{tabular}
\caption{Benchmark point BP2. Luminosity $\mathcal L = 100$ fb$^{-1}$.
Universal acceptance cuts = $|\eta_l| < 2.5$, $|\eta_j| < 3$, $p^T_{j_{1,2}} > 30$ GeV,  $\Delta R_{jj} > 0.4$ + $\Delta R_{lj} > 0.4$ + $\Delta R_{ll} > 0.3$. Transverse momentum cuts for the $eee$ category: 
$p^T_{e_{1, 2}} > 28$ GeV and $p^T_{e_{3}} > 5$ GeV. Transverse momentum cuts for the $ee\mu$ category: $p^T_{e_{1, 2}} > 14$ GeV and $p^T_{\mu} > 10$ GeV. }
\label{fig.H2BP2.1}
\end{sidewaystable}

\begin{sidewaystable}[t]
\begin{tabular}{|c||c|c||c|c||c|c||c|c||c|} \hline
				& BP2($e\mu\mu$)	& Eff. \%	& $WZjj$	& Eff. \%	& $t \bar t l \nu$	& Eff. \%	& $t \bar t$	& Eff. \% 	& $S/\sqrt{B}$			\\ \hline \hline
No cuts.			 & 15.8573	 & 100			 & 19613.8	 & 100			 & 187.639	 & 100			 & 21109.9	 & 100	& 0.0783984 \\
$\eta$			 & 12.1841	& 76.84			 & 10031.1	& 51.14			 & 132.82	& 70.78			 & 16298.2	& 77.21	& 0.0748997 \\
$p^{T}$			 & 3.99205	& 32.76			 & 3776.77	& 37.65			 & 81.6147	& 61.45			 & 3599.54	& 22.09	& 0.046226 \\
$\Delta R$			 & 3.35443	& 84.03			 & 3287.72	& 87.05			 & 69.8418	& 85.58			 & 0	& 0	& 0.0578904 \\
$|M_{jj} - M_W| < 20$ GeV			 & 3.28512	& 97.93			 & 394.687	& 12			 & 9.82872	& 14.07			 & 0	& 100	& 0.163337 \\
$|M_{l^+l^-} - M_Z| > 20$ GeV			 & 2.48117	& 75.53			 & 25.6999	& 6.511			 & 8.02204	& 81.62			 & 0	& 100	& 0.427268 \\
\hline \hline
$|M^T_{vis} - m_{H_{2}}| < 50$ GeV			 & 2.3079	& 93.02			 & 6.11015	& 23.77			 & 1.48265	& 18.48			 & 0	& 100	& 0.837561 \\
$|M_{all} - m_{H_{2}}| < 50$ GeV			 & 2.06533	& 83.24			 & 4.24445	& 16.52			 & 0.864063	& 10.77			 & 0	& 100	& 0.913781 \\
\hline
\end{tabular}
\\
\bigskip \bigskip \bigskip \\
\begin{tabular}{|c||c|c||c|c||c|c||c|c||c|} \hline
				& BP2($\mu\mu\mu$)	& Eff. \%	& $WZjj$	& Eff. \%	& $t \bar t l \nu$	& Eff. \%	& $t \bar t$	& Eff. \% 	& $S/\sqrt{B}$			\\ \hline \hline
No cuts.			 & 14.3326	 & 100			 & 18102	 & 100			 & 64.9422	 & 100			 & 6901.4	 & 100	& 0.0905235 \\
$\eta$			 & 11.0266	& 76.93			 & 9214.47	& 50.9			 & 45.9328	& 70.73			 & 5336.08	& 77.32	& 0.0912682 \\
$p^{T}$			 & 4.33858	& 39.35			 & 3910.77	& 42.44			 & 32.4907	& 70.74			 & 1428.14	& 26.76	& 0.0591975 \\
$\Delta R$			 & 3.6871	& 84.98			 & 3401.86	& 86.99			 & 27.7482	& 85.4			 & 0	& 0	& 0.0629597 \\
$|M_{jj} - M_W| < 20$ GeV			 & 3.62472	& 98.31			 & 413.624	& 12.16			 & 3.85883	& 13.91			 & 0	& 100	& 0.177401 \\
$|M_{l^+l^-} - M_Z| > 20$ GeV			 & 1.54553	& 42.64			 & 15.252	& 3.687			 & 2.16016	& 55.98			 & 0	& 100	& 0.370383 \\
\hline \hline
$|M^T_{vis} - m_{H_{2}}| < 50$ GeV			 & 1.35147	& 87.44			 & 3.21832	& 21.1			 & 0.422213	& 19.55			 & 0	& 100	& 0.708313 \\
$|M_{all} - m_{H_{2}}|  < 50$ GeV		 & 1.18514	& 76.68			 & 2.19219	& 14.37			 & 0.186559	& 8.636			 & 0	& 100	& 0.768414 \\
\hline
\end{tabular}
\caption{Benchmark point BP2. Luminosity $\mathcal L = 100$ fb$^{-1}$.
Universal acceptance cuts = $|\eta_l| < 2.5$, $|\eta_j| < 3$, $p^T_{j_{1,2}} > 30$ GeV,  $\Delta R_{jj} > 0.4$ + $\Delta R_{lj} > 0.4$ + $\Delta R_{ll} > 0.3$. Transverse momentum cuts for the $eee$ category: 
$p^T_{e_{1, 2}} > 35$ GeV and $p^T_{e_{3}} > 5$ GeV. Transverse momentum cuts for the $ee\mu$ category: $p^T_{e_{1, 2}} > 14$ GeV and $p^T_{\mu} > 10$ GeV. }
\label{fig.H2BP2.2}
\end{sidewaystable}

\subsection{The heavy $Z'$ gauge boson mediator}

In this section, we analyse the channel 
$pp\rightarrow Z' \rightarrow \nu_h\nu_h\rightarrow 3l + 2j + E_{T}^{\rm miss}$. This process has been extensively studied in the past, see for example Refs.~\cite{Basso:2008iv}, as it was considered the default production mechanism of heavy neutrinos. Owing to the present bounds on the mass of the extra heavy $Z'$ boson, the cross section is now expected to be quite small. As shown in Fig.~\ref{Z'cross-section}, $\sigma_{pp\rightarrow Z'\rightarrow\nu_h\nu_h}\le 0.5$ fb for $M_{Z'}\ge 4$ TeV. 

The scope of our analysis is investigating new techniques in the attempt to enhance the heavy neutrino pair production signal over the background. As the $Z'$ boson is quite massive, the produced heavy neutrinos are likely to be highly boosted for masses $m_{\nu_h}\le$ 500 GeV. Their decay products will be boosted as well, giving rise to very collimated configurations in the final state. For higher masses $m_{\nu_h}\ge$ 500 GeV, the neutrinos wouldn't be boosted anymore but the two $W$-bosons coming from their decay would be. Therefore in either case, the two jets coming from the semi-leptonic decay of the neutrino, $\nu_h\rightarrow ljj$, will very likely be so close each other to form a unique fat jet. This distinctive feature can be exploited in order to disentangle the signal from the overwhelming SM background. What differentiates the two cases of a relatively light and a quite heavy neutrino is the sub-structure of such a fat jet. When the heavy neutrino is relatively light and boosted, the fat jet will include inside its cone the lepton coming from the same three-body decay. If the neutrino is quite heavy, the lepton will instead remain well separated from the fat jet. 

In order to depict the two highlighted situations, we choose the following set of parameters allowed by present direct and indirect searches: $M_{Z'} = 4$ TeV, $g' = 0.3$ and $\tilde g = -0.3$ with a decoupled heavy Higgs. We consider the two benchmark points characterised by different heavy neutrino masses: BP1 with $m_{\nu_h} = 400$ GeV ($\sigma (pp\rightarrow Z'\rightarrow \nu_h\nu_h$ = 0.37 fb) and BP2 with $m_{\nu_h} =1000$ GeV ($\sigma (pp\rightarrow Z'\rightarrow \nu_h\nu_h$ = 0.26 fb) . 

\begin{figure}[t]
\centering
\subfigure[most energetic lepton $p^T$]{\includegraphics[scale=0.37]{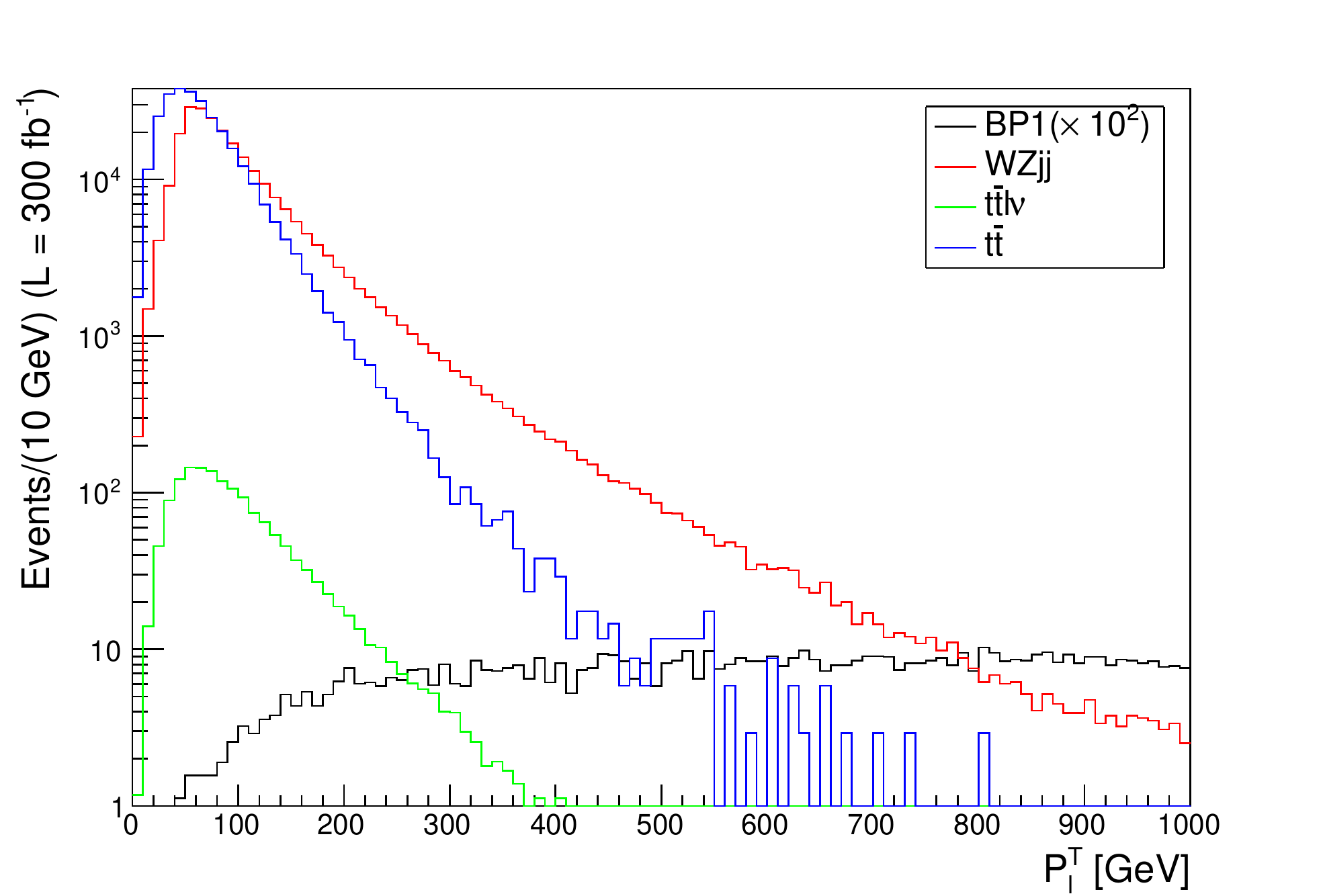}} 
\subfigure[second most energetic lepton $p^T$]{\includegraphics[scale=0.37]{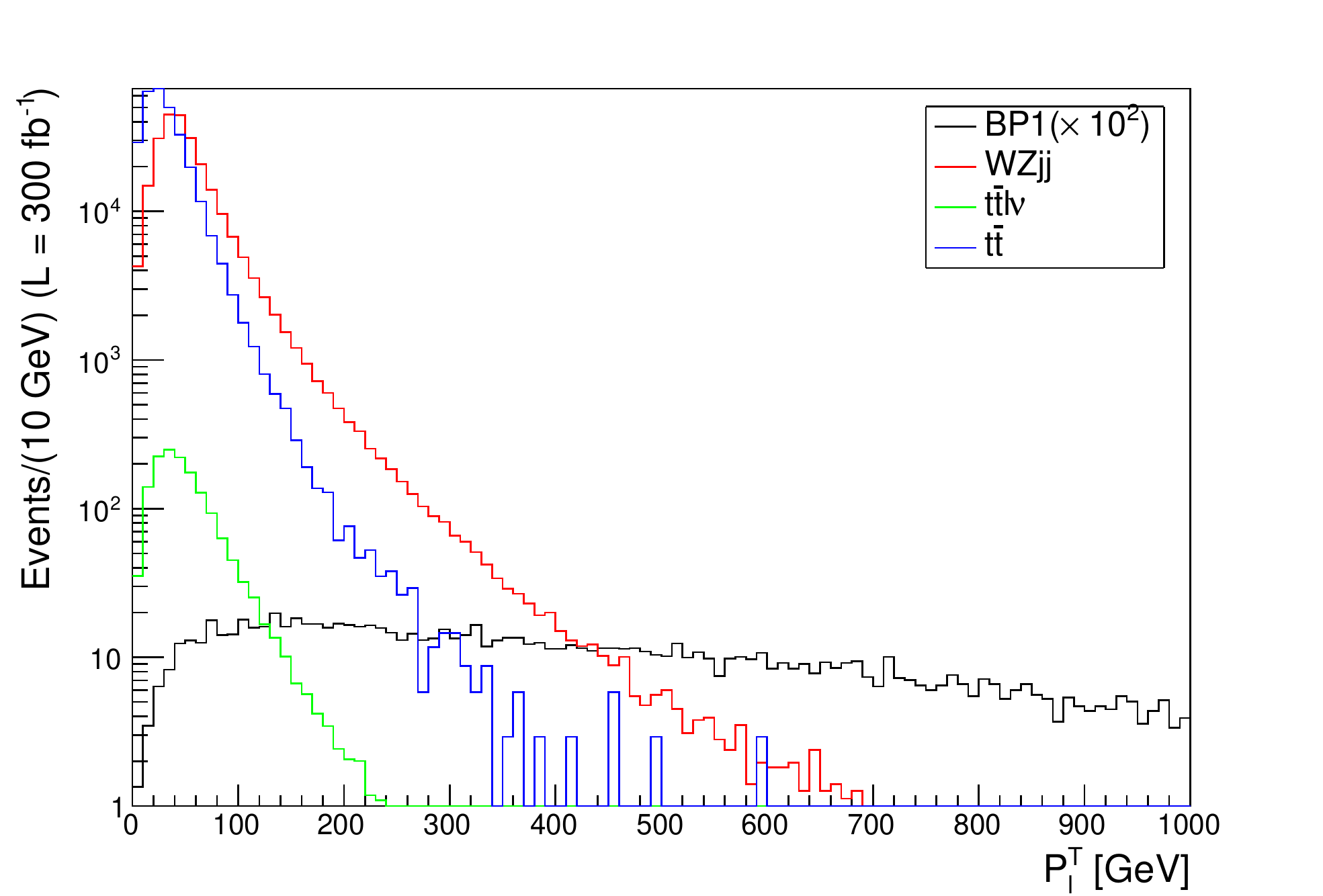}} 
\subfigure[least energetic lepton $p^T$]{\includegraphics[scale=0.37]{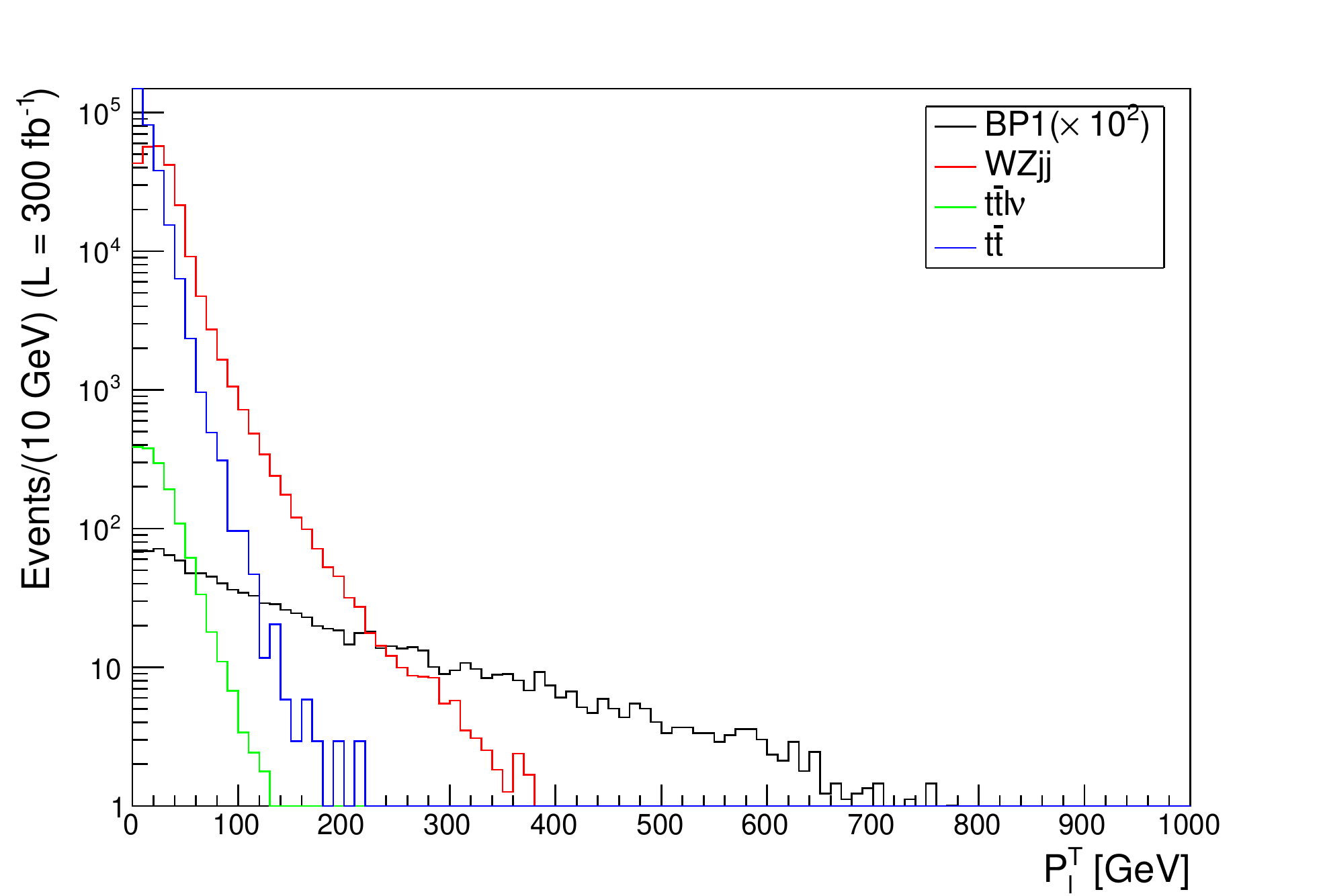}} 
\subfigure[leading and sub leading jet $p^T$]{\includegraphics[scale=0.37]{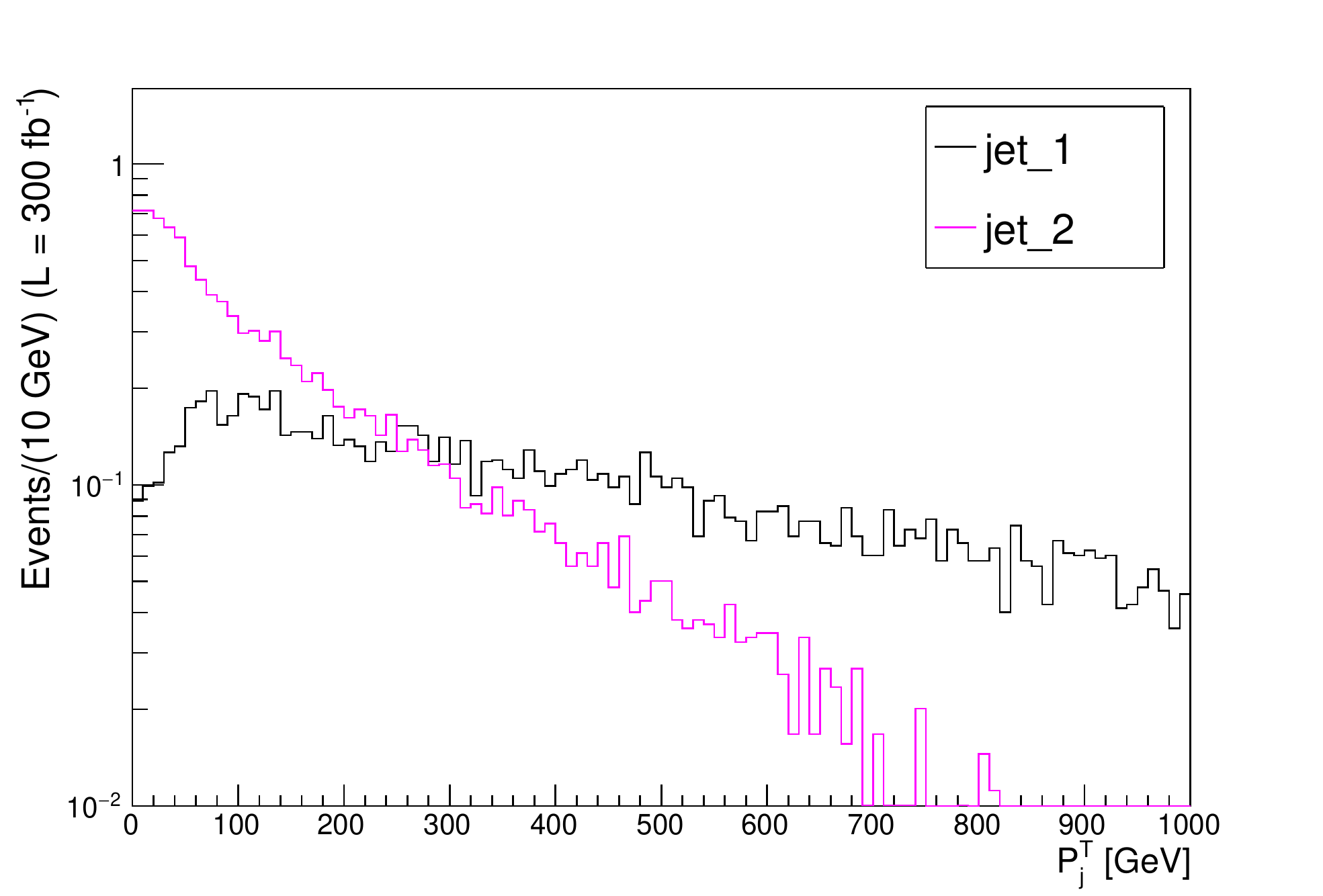}}
\subfigure[lepton - jet angular separation $\Delta R_{lj}$]{\includegraphics[scale=0.37]{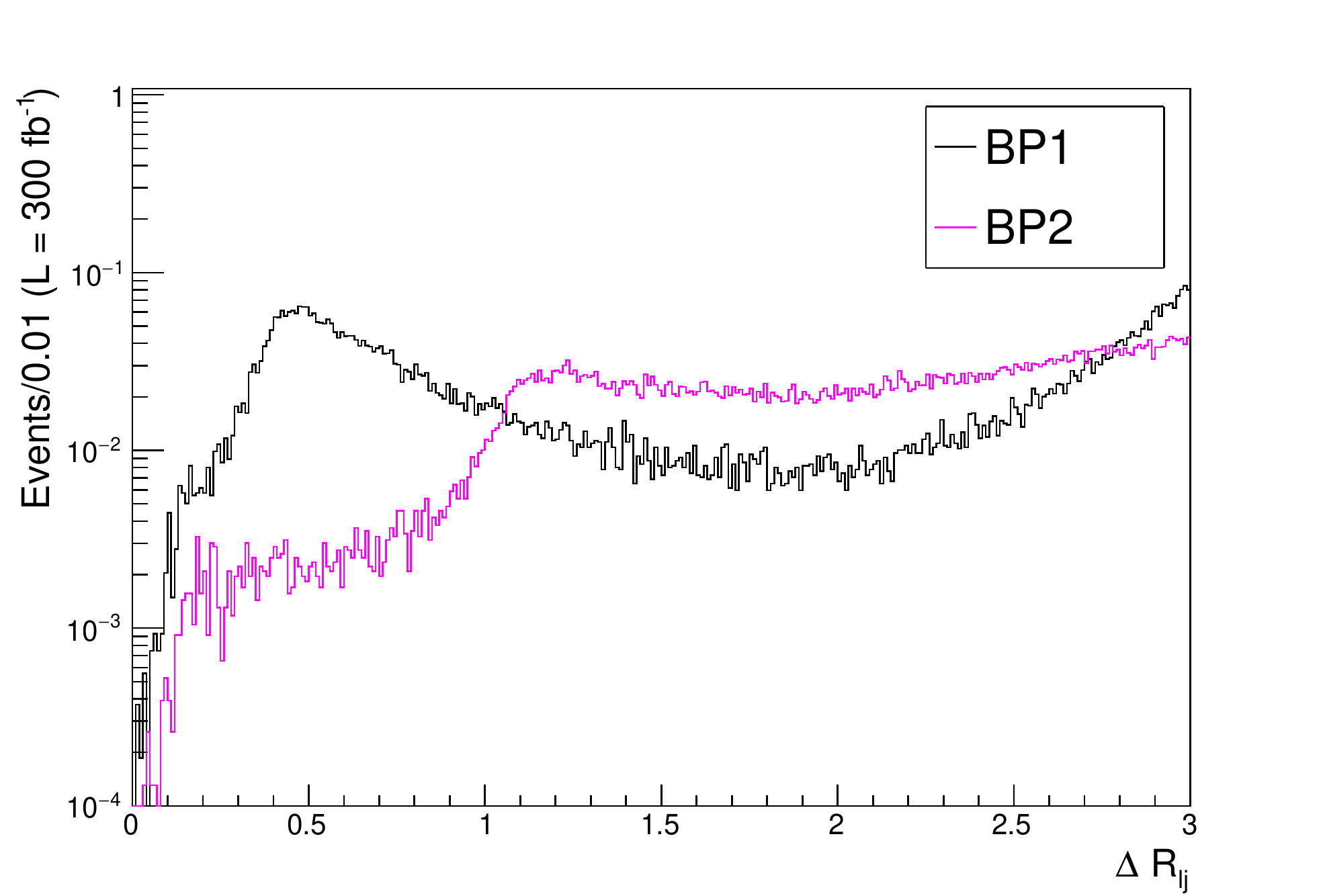}} 
\subfigure[jet angular separation $\Delta R_{jj}$]{\includegraphics[scale=0.37]{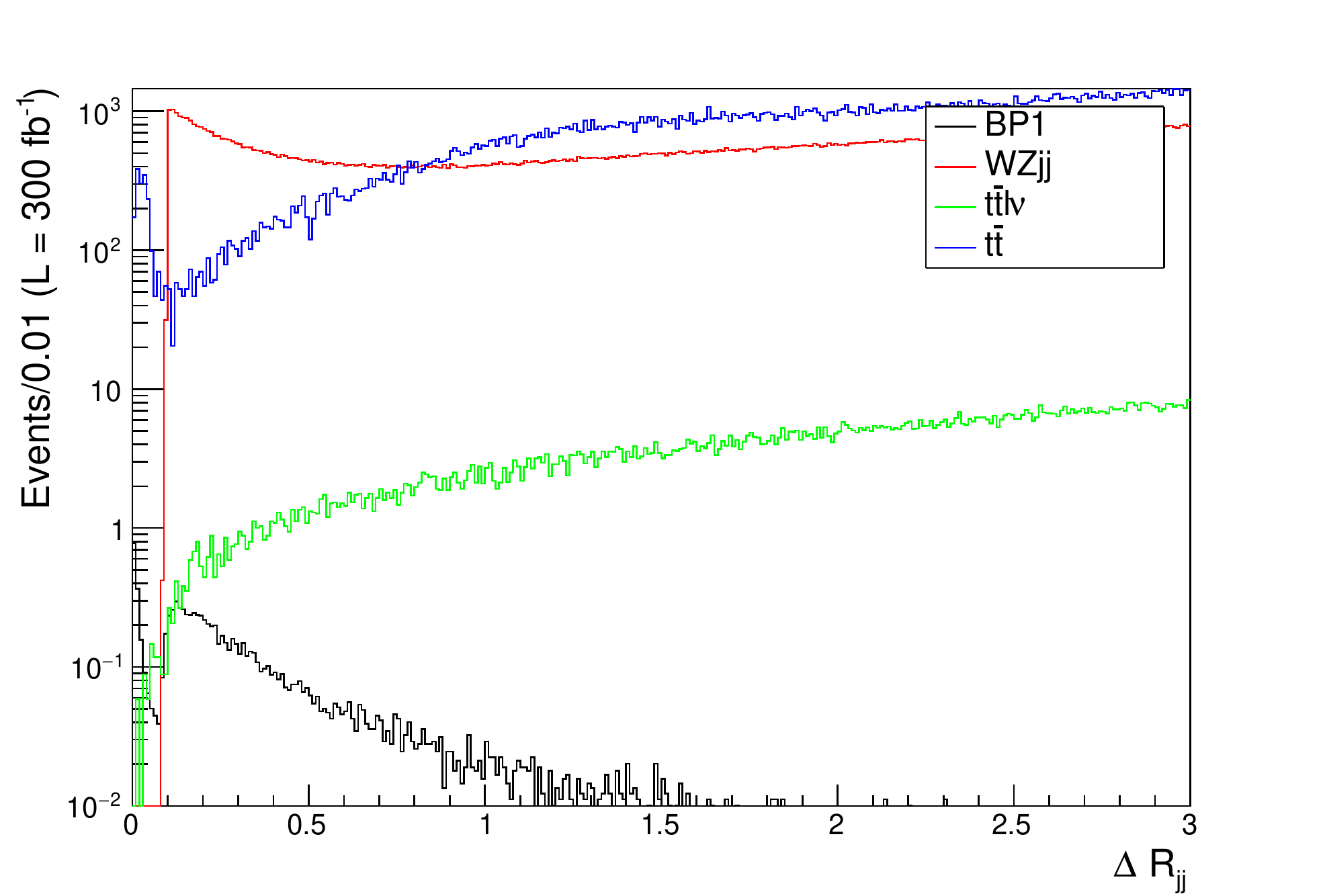}}
\caption{(a) Distribution in the transverse momentum of the leading lepton. The black solid line is the signal, the red line the $WZjj$ background, the green and the blue lines the $t\bar t l\nu$ and $t\bar t$ SM background components. (b) same as (a) for the second most energetic lepton. (c) same as (a) for the least energetic lepton. (d) Distribution in the transverse momentum of the two jets in BP1. Black and magenta lines refer to the leading and sub-leading jet, respectively. (e) Distribution in the angular separation between any jet and lepton averaged over the six possible pairs. The black and magenta lines refer to BP1 and BP2, respectively. (f) Distribution in the angular separation between the two jets. The color code is the same as for plot (a). All distributions are calculated for the benchmark point BP1 (except plot (e)) for a luminosity $\mathcal L = 300$ fb$^{-1}$.
\label{fig.BP1-Pt-DR}}
\end{figure}

To start with, we analyse the kinematics of the produced $3l + 2j + E_{T}^{\rm miss}$ events.  We firstly  examine the BP1 scenario, characterised by the production of a pair of boosted heavy neutrinos. In Fig.~\ref{fig.BP1-Pt-DR}, we show the transverse momentum distribution of the leading lepton (plot a), the sub-leading lepton (plot b), the third less energetic lepton (plot c) and of the two jets (plot d). In estimating the signal and background efficiencies, unlike the previous $H_2$-mediated process, we show the sum of the events over the four lepton categories. Indeed, there is no much difference in the efficiency between the four signal types, $ee\mu$, $e\mu\mu$ and $\mu\mu\mu$ , owing to the much larger heavy neutrino mass. Altogether, the basic selection cuts have an average efficiency of about 75\%. In order to suppress the SM background, we impose more stringent cuts on the transverse momentum of the leptons. Based on the observation of plots (a) - (c) of Fig.~\ref{fig.BP1-Pt-DR}, the following constraints have been chosen:

\begin{itemize}
\item $p^T_{l_{1}} > 300$ GeV, $p^T_{l_{2}} > 150$ GeV, $p^T_{l_{3}} > 15$ GeV
\end{itemize}

The efficiency of the signal remains 64\% while the background sources are heavily suppressed, with an efficiency of less than 1\%.  In Fig.~\ref{fig.BP1-Pt-DR}, we show the distribution in $\Delta R_{lj}$ (plot e), averaged over the six possible lepton-jet pairs, and $\Delta R_{jj}$ (plot f). Clearly, while any leptons and jets are well separated from each other (same is true for the separation between any two leptons), the two jets coming from the $W$-boson decay are typically collimated. Requiring the two jets and the three leptons to be resolved separately via $\Delta R_{jj, lj} > 0.4$ and $\Delta R_{ll} > 0.3$ takes the signal efficiency down to roughly 20\%. Globally, these selection cuts reduce the signal by a factor of seven but they help decrease the SM background substantially, which goes down by almost three orders of magnitude. Notice that the separation lepton-jet suppresses completely the $t\bar t$ source of background. The additional cut $\Delta R_{jj} < 1.5$ further reduces the $WZjj$ and $t\bar t l\nu$ background components by two thirds. Finally, two cuts on the same-flavour opposite-sign di-lepton and di-jet invariant masses are imposed. The global efficiencies are shown in Tab.~\ref{fig.Z'BP1}. Despite of all these cuts, the SM background is still of the same order as the signal. Looking at the total visible mass distribution of the final state, one realises that signal and background occupy the same regions. The cut $|M^T_{vis} - M_{Z'}|\le$ 1 TeV reduces the SM background to zero, affecting the signal only marginally. We obtain an analogous result if we impose $|M_{all} - M_{Z'}|\le$ 1 TeV via the light neutrino reconstruction algorithm. Despite the success of suppressing the huge SM background, the number of signal events left is very small (one in this specific case). 

\begin{figure}[t]
\centering
\subfigure[]{\includegraphics[scale=0.35]{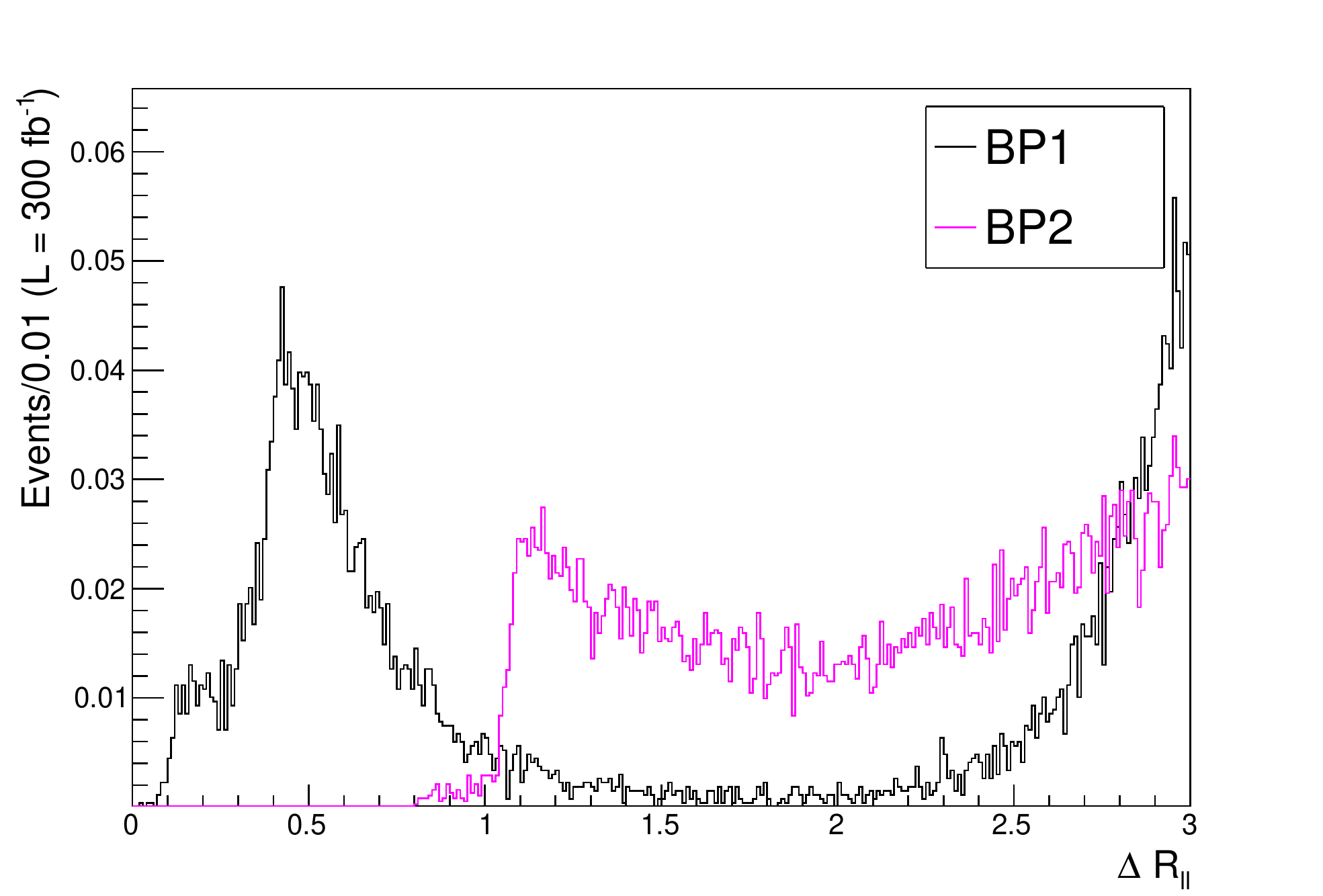}} 
\subfigure[]{\includegraphics[scale=0.35]{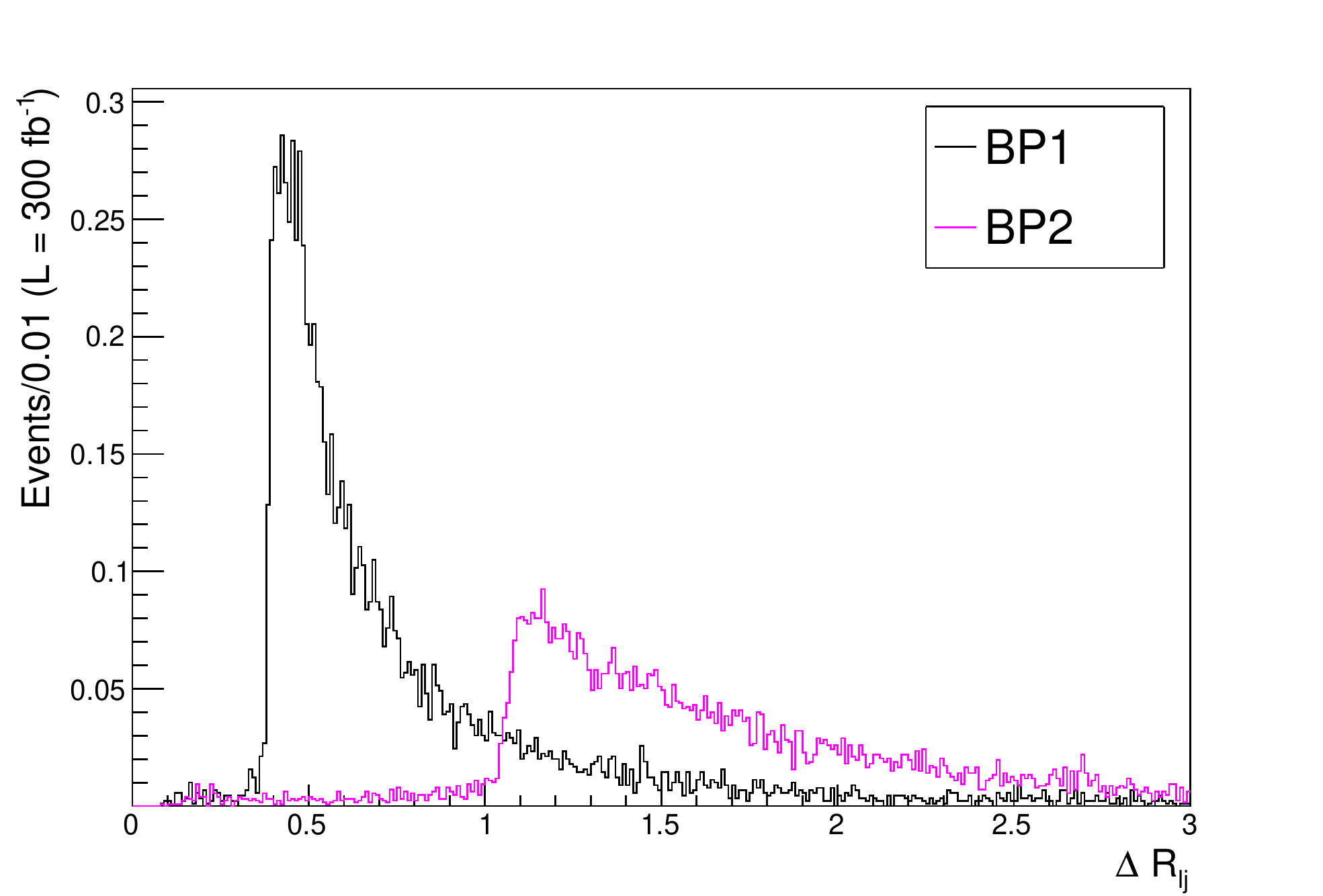}}
\caption{(a) Distribution in the angular separation between the leptons averaged over the three possible pairs. Black and magenta lines refer to BP1 and BP2 scenarios, respectively. (b) Distribution in the angular separation between the fat jet and the closest lepton. The color code is the same as in (a). A luminosity $\mathcal L = 300$ fb$^{-1}$ is assumed.
\label{fig.BP1-DRclosest}}
\end{figure}

Similar conclusions hold for the benchmark point BP2 where $m_{\nu h}$ = 1 TeV. The distinctive characteristic compared to the previous case is the angular separation of the leptons from the jets. As one can see in Fig.~\ref{fig.BP1-Pt-DR}(e), the $\Delta R_{lj}$ is shifted towards higher values. Therefore, the corresponding signal efficiencies increase. As a result, despite the lower initial cross section, the number of events after the cut flow is roughly the same as shown in Tab.~\ref{fig.Z'BP1}. So our conclusion is that this analysis, carried out with standard techniques, requires very high luminosity and is thus tailored for the HL-LHC. In the next section, we explore an advanced experimental strategy based on the so called fat jet. This attempt relies on the observation that the two produced jets are quite collimated, giving rise to a unique large hadronic cone, owing either to the boosted heavy neutrino (BP1) or to the boosted $W$-boson coming from the heavy neutrino decay (BP2).

\begin{sidewaystable}[t]
\begin{tabular}{|c||c|c||c|c||c|c||c|c|c|} \hline
& BP1
& Eff. \% & $WZjj$
& Eff. \% & $t \bar t l \nu$
& Eff. \% & $t \bar t$
& Eff. \% & $S/\sqrt{B}$
\\ \hline \hline
No cuts.			 & 11.4362	 & 100			 & 242279	 & 100			 & 1505.03	 & 100			 & 294324	 & 100			 & 0.01559 \\
$\eta$			 & 10.5155	& 91.95			 & 124514	& 51.39			 & 1062.12	& 70.57			 & 224375	& 76.23			 & 0.019332 \\
$p^{T}$			 & 6.87307	& 65.36			 & 1073.8	& 0.8624			 & 5.39058	& 0.5075			 & 93.3936	& 0.04162			 & 0.307085 \\
$\Delta R$ cuts			 & 1.39939	& 20.36			 & 628.972	& 58.57			 & 3.26969	& 60.66			 & 0	& 0			 & 0.273344 \\
$\Delta R_{jj} < 1.5$			 & 1.27106	& 90.83			 & 178.267	& 28.34			 & 0.412394	& 12.61			 & 0	& 100			 & 0.104689 \\
$|M_{jj} - M_W| < 20$ GeV			 & 1.24427	& 97.89			 & 35.8213	& 20.09			 & 0.117827	& 28.57			 & 0	& 100			 & 0.212022 \\
$|M_{l^+l^-} - M_Z| > 20$ GeV			 & 1.14049	& 91.66			 & 0.699635	& 1.953			 & 0.117827	& 100			 & 0	& 100			 & 1.3762 \\
$|M^T_{vis} - M_{Z'}| < 1$ TeV			 & 0.950781	& 83.37			 & 0	& 0			 & 0	& 0			 & 0	& 100			 & - \\
\hline
\end{tabular}
\\
\bigskip \bigskip \bigskip \\
\begin{tabular}{|c||c|c||c|c||c|c||c|c|c|} \hline
& BP2
& Eff. \% & $WZjj$
& Eff. \% & $t \bar t l \nu$
& Eff. \% & $t \bar t$
& Eff. \% & $S/\sqrt{B}$
\\ \hline \hline
No cuts.			 & 7.62627	 & 100			 & 242279	 & 100			 & 1505.03	 & 100			 & 294324	 & 100			 & 0.0103963 \\
$\eta$			 & 7.04229	& 92.34			 & 124514	& 51.39			 & 1062.12	& 70.57			 & 224375	& 76.23			 & 0.0128916 \\
$p^{T}$			 & 5.04656	& 71.66			 & 1073.8	& 0.8624			 & 5.39058	& 0.5075			 & 93.3936	& 0.04162			 & 0.205656 \\
$\Delta R$ cuts			 & 1.33571	& 26.47			 & 628.972	& 58.57			 & 3.26969	& 60.66			 & 0	& 0			 & 0.200703 \\
$\Delta R_{jj} < 1.5$			 & 1.30044	& 97.36			 & 178.267	& 28.34			 & 0.412394	& 12.61			 & 0	& 100			 & 0.0999255 \\
$|M_{jj} - M_W| < 20$ GeV			 & 1.27536	& 98.07			 & 35.8213	& 20.09			 & 0.117827	& 28.57			 & 0	& 100			 & 0.216923 \\
$|M_{l^+l^-} - M_Z| > 20$ GeV			 & 1.18835	& 93.18			 & 0.699635	& 1.953			 & 0.117827	& 100			 & 0	& 100			 & 1.41058 \\
$|M^T_{vis} - M_{Z'}| < 1$ TeV			 & 1.10055	& 92.61			 & 0	& 0			 & 0	& 0			 & 0	& 100			 & inf \\
\hline
\end{tabular}
\caption{Benchmark point BP1 for the $Z'$-mediator case. Luminosity $\mathcal L = 300$ fb$^{-1}$.
Universal acceptance cuts = $|\eta_l| < 2.5$, $|\eta_j| < 3$, $p^T_{j_{1,2}} > 30$ GeV,  $\Delta R$ cuts = $\Delta R_{jj} > 0.4$ + $\Delta R_{lj} > 0.4$ + $\Delta R_{ll} > 0.3$. Transverse momentum cuts: 
$p^T_{l_{1}} > 300$ GeV, $p^T_{l_{2}} > 150$ GeV and $p^T_{l_{3}} > 15$ GeV. }
\label{fig.Z'BP1}
\end{sidewaystable}

\subsubsection{Fat jet technique}

In this section, we exploit the fat jet techniques to see whether the significance of the $Z'$-mediated production of heavy neutrinos can be increased. With respect to the previous case, where the two jets are isolated and separately reconstructed, one is forced to introduce $WZj$ in the simulation as another source of background. \\
The fat jet is defined by the following two categories:

\begin{itemize}
\item one single leading jet with $p^T_{j_{1}} > 200$ GeV 
\item two jets with separation $\Delta R_{jj}\le 0.8$ and global transverse momentum $p^T_J\ge$ 200 GeV, with $J = j_1 + j_2$.
\end{itemize}

As displayed in Tab.~\ref{fig.Fatjet}, the selection of a fat jet has an intrinsic signal efficiency of more than 79\%, depending on the transverse momentum threshold and the mass of the heavy neutrino. This is a significant improvement as the default di-jet selection, which we discussed previously, has only an efficiency of less than 20\% for the signal. The choice of the $P^T$ threshold for identifying the fat jet is not really significant for the signal. In the range $100 \, \textrm{GeV} \le p^T_J\le 200$ GeV, no dramatic variation is observed in the selection efficiency as shown in Tab.~\ref{fig.Fatjet}. Where it really helps is in reducing the SM background. For that reason, we choose the value $p^T_J > 200$ GeV.  We also impose that all the sub-leading jets, if present, have $p^T_j < 0.25 \, p^T_J$. This restriction has the advantage to clean the signal (in which case the two jets are merged into a single fat jet) removing any other jet activity originating from initial state radiation, pile-up or $\tau$ hadronic decays. It also suppresses the backgrounds (with the only exception of $WZj$) as the two jets are expected to be well isolated from each other. Therefore, requiring a hierarchy between the leading fat jet and the sub leading one will reject all the events for which the two jets are characterised by the same $p^T$. \\
For the leptons, besides the standard geometric acceptance requirement $|\eta_l| < 2.5$, we apply the set of cuts on the transverse momentum as before: $p^T_{l_{1}} > 300$ GeV, $p^T_{l_{2}} > 150$ GeV and $p^T_{l_{3}} > 15$ GeV. This selection is particularly effective in suppressing the background to a few percent of its initial value. The selections described above are common to both the two BPs and rely only on the boosted topology and on the very energetic nature of the final state objects, which is a direct consequence of the large $Z'$ mass. 

For the BP1 we furthermore require that one lepton lies inside the fat jet, while the other two leptons are well separated one from each other via $\Delta R_{ll} > 0.3$ and outside the fat jet cone, $\Delta R_{lJ} > 0.8$. This choice is motivated by the kinematic features of the BP, described in the previous section, and supported by the analysis of the angular separation between these particles. \\
In Fig.~\ref{fig.BP1-DRclosest}, we plot the distribution in the separation variable between two leptons $\Delta R_{ll}$, averaged over all lepton pairs, (plot a) and between the fat jet and its closest lepton $\Delta R_{lJ}$ (plot b). From there, it is evident that one of the leptons is very close to this fat jet. If we require that the three leptons are all well separated from the fat jet,  $\Delta R_{lJ}\ > 0.8$, the number of signal events drops by a factor of five. The most probable configuration is when the lepton is inside the fat. When we require two leptons to be separated from the fat jet, the efficiency in fact increases up to about 82\% as shown in the sixth row of Tab.~\ref{fig.Fatjet}. \\
In principle, the lepton that may fall inside the fat jet can be disentangled from the jet and individually reconstructed. An additional cut that is particularly effective in  suppressing the background while leaving the signal almost unaffected consists of imposing a lower value on the transverse momentum of the lepton with respect to the direction of the jet \cite{Sirunyan:2017uhk}. In particular, this cut removes the SM $t\bar t$ background and compensates for the absence of the isolation requirement on the lepton inside the fat jet which was very effective in the standard analysis in the removal of events with a lepton emerging from the $B$ meson decay. \\
Finally, the cut on the invariant mass of lepton pairs is imposed as described in the previous analyses in order to reject events from the $WZjj$ and $WZj$ backgrounds where two same flavour and opposite sign leptons are generated from the $Z$ decay. \\
The full cut flow is shown in Tab.~\ref{fig.Fatjet2} for the two benchmark points BP1 and BP2. After the full cut flow, the SM background is zero. The number of signal events is around four at $\mathcal L = 300$ fb$^{-1}$. Even though this means a small number of events given the projected luminosity of Run II, such a number can become significant at the HL-LHC stage, as the signal is almost background free. 

Analogous results are obtained for the benchmark point BP2. This scenario, characterised by a very heavy neutrino $m_{\nu h} =$ 1000 GeV, differs from the previous one in the angular separation between leptons and fat jet. In this case all the leptons are well separated from the fat jet ($\Delta R_{lJ} > 0.8$) as clear from Fig.~\ref{fig.BP1-DRclosest}(a).
 We therefore do not require the presence of any lepton inside the fat jet. The remaining cut flow is identical to the BP1 scenario and at the end the number of signal events is same, i.e. four at $\mathcal L = 300$ fb$^{-1}$. These events are also almost background free as shown in Tab.~\ref{fig.Fatjet2}.

The fat jet analysis appears to be more competitive than the standard technique in searching for these low signal yields. For the discussed characteristics, this seems to be an ideal study for the HL-LHC option \cite{Gianotti:2002xx}, which goes beyond this particular $Z'$-boson analysis.

\begin{sidewaystable}[t]
\begin{tabular}{|c||c|c||c|c||c|c||c|c||c|c|c|} \hline
& BP1
& Eff. \% & $WZjj$
& Eff. \% & $t \bar t l \nu$
& Eff. \% & $t \bar t$
& Eff. \% & $WZj$
& Eff. \% \\ \hline \hline
Fat Jet(100)  & 10.1885 & 89.09
& 113051 & 46.66
& 765.197 & 50.84
& 99452.5 & 27.38
& 47731 & 24.8
\\
Fat Jet(130)  & 9.70868 & 84.89
& 85950.7 & 35.48
& 467.007 & 31.03
& 48839 & 13.45
& 33767.8 & 17.55
\\
Fat Jet(150)  & 9.46763 & 82.79
& 71715.2 & 29.6
& 328.855 & 21.85
& 31891 & 8.781
& 27100.2 & 14.08
\\
Fat Jet(170)  & 9.29913 & 81.31
& 59914.6 & 24.73
& 231.353 & 15.37
& 22300.6 & 6.14
& 21814.6 & 11.34
\\
Fat Jet(200)  & 9.05139 & 79.15
& 45985.6 & 18.98
& 136.296 & 9.056
& 14429.3 & 3.973
& 15986 & 8.307
\\
\hline
\end{tabular}
\\
\bigskip \bigskip \bigskip \\
\begin{tabular}{|c||c|c||c|c||c|c||c|c||c|c|c|} \hline
& BP2
& Eff. \% & $WZjj$
& Eff. \% & $t \bar t l \nu$
& Eff. \% & $t \bar t$
& Eff. \% & $WZj$
& Eff. \% \\ \hline \hline
Fat Jet(100)  & 7.12303 & 93.4
& 113051 & 46.66
& 765.197 & 50.84
& 99452.5 & 27.38
& 47731 & 24.8
\\
Fat Jet(130)  & 6.86827 & 90.06
& 85950.7 & 35.48
& 467.007 & 31.03
& 48839 & 13.45
& 33767.8 & 17.55
\\
Fat Jet(150)  & 6.73266 & 88.28
& 71715.2 & 29.6
& 328.855 & 21.85
& 31891 & 8.781
& 27100.2 & 14.08
\\
Fat Jet(170)  & 6.58529 & 86.35
& 59914.6 & 24.73
& 231.353 & 15.37
& 22300.6 & 6.14
& 21814.6 & 11.34
\\
Fat Jet(200)  & 6.42773 & 84.28
& 45985.6 & 18.98
& 136.296 & 9.056
& 14429.3 & 3.973
& 15986 & 8.307
\\
\hline
\end{tabular}
\caption{BP1 and BP2 benchmark points for the $Z'$-mediator case. Luminosity $\mathcal L = 300$ fb$^{-1}$. Fat jet definition (first column) and efficiencies for signal and SM background as explained in the text. The minimum $p^T$ in GeV required for the fat jet is given in parenthesis. The efficiencies are not cumulative.}
\label{fig.Fatjet}
\end{sidewaystable}

\begin{sidewaystable}[t]
\begin{tabular}{|c||c|c||c|c||c|c||c|c||c|c|c|c|} \hline
& BP1
& Eff. \% & $WZjj$
& Eff. \% & $t \bar t l \nu$
& Eff. \% & $t \bar t$
& Eff. \% & $WZj$
& Eff. \% & $S/\sqrt{B}$
\\ \hline \hline
No cuts.  & 11.4362 & 100
& 242279 & 100
& 1505.03 & 100
& 363181 & 100
& 192436 & 100
& 0.0122192 \\
Fat Jet(200)  & 9.05139 & 79.15
& 45985.6 & 18.98
& 136.296 & 9.056
& 14429.3 & 3.973
& 15986 & 8.307
& 0.0413375 \\
$p^T_j/p^T_J < 0.25$ 
& 9.04804 & 99.96
& 31324.6 & 68.12
& 43.7137 & 32.07
& 8974.54 & 62.2
& 15986 & 100
& 0.0381373 \\
$|\eta_{l}| < 2.5$
& 8.60278 & 95.08
& 24281 & 77.51
& 38.0875 & 87.13
& 7964.72 & 88.75
& 12118.1 & 75.8
& 0.0429392 \\
$p^{T}_{l_i} > 300,150,15$ GeV
& 6.74028 & 78.35
& 1046.65 & 4.311
& 1.14881 & 3.016
& 70.0452 & 0.8794
& 392.486 & 3.239
& 0.221362 \\
$1l$ inside Fat Jet
& 5.54734 & 82.3
& 131.252 & 12.54
& 0.972071 & 84.62
& 64.2081 & 91.67
& 57.307 & 14.6
& 0.42314 \\
$\Delta R_{ll} > 0.3$
& 4.92353 & 88.75
& 95.85 & 73.03
& 0.913158 & 93.94
& 55.4524 & 86.36
& 43.9798 & 76.74
& 0.396041 \\
$p^{T,rel}_l > 200$ GeV
& 4.0888 & 83.05
& 3.91796 & 4.088
& 0.0294567 & 3.226
& 0 & 0
& 1.33272 & 3.03
& 2.14266 \\
$|M_{l^+l^-} - M_Z| > 20$ GeV
& 3.75737 & 91.89
& 0.139927 & 3.571
& 0.0294567 & 100
& 0 & 100
& 0 & 0
& 9.93483 \\
\hline
\end{tabular}
\\
\bigskip \bigskip \bigskip \\
\begin{tabular}{|c||c|c||c|c||c|c||c|c||c|c|c|c|} \hline
& BP2
& Eff. \% & $WZjj$
& Eff. \% & $t \bar t l \nu$
& Eff. \% & $t \bar t$
& Eff. \% & $WZj$
& Eff. \% & $S/\sqrt{B}$
\\ \hline \hline
No cuts.  & 7.62627 & 100
& 242279 & 100
& 1505.03 & 100
& 363181 & 100
& 192436 & 100
& 0.00852962 \\
Fat Jet(200)  & 6.42773 & 84.28
& 45985.6 & 18.98
& 136.296 & 9.056
& 14429.3 & 3.973
& 15986 & 8.307
& 0.0275661 \\
$p^T_j/p^T_J < 0.25$
& 6.42773 & 100
& 31324.6 & 68.12
& 43.7137 & 32.07
& 8974.54 & 62.2
& 15986 & 100
& 0.0270827 \\
$|\eta_{l}| < 2.5$
& 6.00288 & 93.39
& 24281 & 77.51
& 38.0875 & 87.13
& 7964.72 & 88.75
& 12118.1 & 75.8
& 0.030504 \\
$p^{T}_{l_i} > 300,150,15$ GeV
& 5.07399 & 84.53
& 1046.65 & 4.311
& 1.14881 & 3.016
& 70.0452 & 0.8794
& 392.486 & 3.239
& 0.154462 \\
$\Delta R_{lJ} > 0.8$
& 4.93289 & 97.22
& 891.195 & 85.15
& 0.117827 & 10.26
& 0 & 0
& 325.184 & 82.85
& 0.145477 \\
$\Delta R_{ll} > 0.3$
& 4.58564 & 92.96
& 637.927 & 71.58
& 0.117827 & 100
& 0 & 100
& 262.879 & 80.84
& 0.164345 \\
$p^{T,rel}_l > 400$ GeV
& 3.36829   & 73.45                  
& 76.5401      & 12                     
& 0    & 0                      
& 0    & 100           
 & 31.3189      & 11.91                  
 & 0.441542 \\
$|M_{l^+l^-} - M_Z| > 20$ GeV
& 3.17075      & 94.14                  
& 1.25934      & 1.645                  
& 0    & 100                    
& 0    & 100                    
& 0.99954 & 3.191                  
& 2.24111 \\
\hline
\end{tabular}
\caption{BP1 and BP2 benchmark points for the $Z'$-mediator case. Luminosity $\mathcal L = 300$ fb$^{-1}$. Cut flow and efficiencies for signal and SM background as explained in the text.}
\label{fig.Fatjet2}
\end{sidewaystable}

\section{Conclusions}
\label{sec:summa}
The experimental evidence of neutrino flavour oscillations, implying that such states of Nature have a mass, is a pressing problem for the SM. An economical solution to account for these is to extend the SM gauge group by an additional Abelian
$U(1)_{B-L}$ symmetry, broken through a Higgs mechanism, thereby yielding simultaneously new Higgs
($H_2$) and gauge ($Z'$) bosons, one of each, acting as heavier companions to the SM(-like) $H_1$ and $Z$ states. As a byproduct of this dynamics, one then gets three additional heavy neutrinos, of Majorana nature, following the customary requirement of avoiding gauge and gravitational anomalies. Then, a Type-I seesaw is responsible for neutrino mass generation and mixing, leaving behind the three SM-like light neutrinos ($\nu_l$) and three heavy ($\nu_h$) ones, which can then be at the EW or TeV scale. Since the limits on additional Higgs and gauge bosons of the kind we introduced here are presently set in the same energy ranges, respectively, one is tempted to access all the additional states of this theoretical construct at once, by pursuing the search for pair production of heavy neutrino states emerging from the decay of $H_2$ and $Z'$ bosons, so that the latter act as portals to the former.

The parameter space of this scenario, tested against all available theoretical and experimental constraints, was already defined in a previous publication \cite{Accomando:2016sge} (see also \cite{Accomando:2016eom,Accomando:2016upc}), where potential feasibility of the aforementioned production and decay modes was highlighted, based on inclusive analyses. In this paper, which follows closely Ref.~\cite{Accomando:2016rpc},
where the $gg\to H_{1,2}\to \nu_h\nu_h$ signal was proved to be accessible during Run 2 of the LHC in essentially a background free environment by exploiting DVs
induced by rather light $\nu_h$ states yielding signatures with  two to four
leptons, we assessed the possibility of accessing heavier $\nu_h$ states, by exploiting $gg\to H_{2}\to \nu_h\nu_h$ 
and $q\bar q\to Z'\to \nu_h\nu_h$  signals, for which DVs are no longer available. The analysis in these conditions is much more challenging, as the heavy neutrino decay products stem from the interaction point, where the SM background is initially 
overwhelming. 

However, upon a dedicated analysis of the $3l+2j+E_T^{\rm miss}$ signature emerging from both production and decay modes (i.e., via $H_2$ and $Z'$ intermediate states), we have been able to prove that some evidence of the existence of $\nu_h$ states can already be glimpsed with standard luminosity conditions during the 13 TeV runs of the LHC, though full discovery will probably have to wait for the high luminosity option of the machine, the so-called High Luminosity LHC
(HL-LHC) \cite{Gianotti:2002xx}. We have come to this encouraging conclusion after adopting a rather sophisticated signal-to-background selection. In doing so, we had to devise different approaches depending on the relative mass differences amongst the $Z'$, $H_2$ and $\nu_h$ states (as well as $W^\pm$ and $Z$ bosons, which appear in the two $\nu_h$ decay chains), since -- depending on these -- one may have more or less boosted objects in the detector, thus ranging from the case of all reconstructed particles being separated to the one where the two jets could be merged by standard jet clsutering algorithms into a single fat one (for which then we had to exploit jet substructure techniques) with and without a lepton inside it. To facilitate our studies, we have made use of a variety of triggers available for current and subsequent runs of the CERN collider.  

As emphasised in    Refs.~\cite{Accomando:2016sge,Accomando:2016eom,Accomando:2016upc}, simultaneous access to
the $gg\to H_1\to \nu_h\nu_h$,
      $gg\to H_2\to \nu_h\nu_h$ and 
  $q\bar q\to Z'\to \nu_h\nu_h$  modes would finally enable one to establish a direct link between measurements obtainable at the EW  scale and the dynamics of the underlying model up to those where a GUT scenario embedding a 
$U(1)_{B-L}$   can be realised.
In short, with the present paper, we have completed the analysis of the  $U(1)_{B-L}$  signatures that need establishing at the LHC in order to pursue such an endeavour. In future publications, we shall see their deployment in fully fledged experimental analyses.

\section*{Acknowledgements}
EA, LDR, SM and CHS-T
are supported in part through the NExT Institute. The work of LDR has been supported
by the STFC/CO\-FUND Rutherford International
Fellowship scheme.

\newpage

\providecommand{\href}[2]{#2}\begingroup\raggedright\endgroup

\end{document}